\documentclass[journal]{IEEEtran}
\usepackage{amsmath}
\usepackage{graphicx}
\usepackage{caption2}
\usepackage{amsthm}
\usepackage{cite}
\begin{document}
\title{DoF Analysis of the MIMO Broadcast Channel with Alternating/Hybrid CSIT}
\author{{Borzoo Rassouli, Chenxi Hao and Bruno Clerckx} %
\thanks{Borzoo Rassouli and Chenxi Hao are with the Communication and Signal Processing group of Department of Electrical and Electronics,
Imperial College London, United Kingdom. emails: b.rassouli12@imperial.ac.uk , chenxi.hao10@imperial.ac.uk}
\thanks{Bruno Clerckx is with the Communication and Signal Processing group of Department of Electrical and Electronics,
Imperial College London and the School of Electrical Engineering, Korea University, Korea. email: b.clerckx@imperial.ac.uk}
\thanks{This paper was presented in part at the IEEE International Conference on Communications (ICC) 2015, London, UK.}
\thanks{This work was partially supported by the Seventh Framework Programme for Research of the European Commission under grant number HARP-318489.}}
\maketitle
\begin{abstract}
%\boldmath
We consider a $K$-user multiple-input single-output (MISO) broadcast channel (BC) where the channel state information (CSI) of user $i(i=1,2,\ldots,K)$ may be instantaneously perfect (P), delayed (D) or not known (N) at the transmitter with probabilities $\lambda_P^i$, $\lambda_D^i$ and $\lambda_N^i$, respectively. In this setting, according to the three possible CSIT for each user, knowledge of the joint CSIT of the $K$ users could have at most $3^K$ states.  In this paper, given the marginal probabilities of CSIT (i.e., $\lambda_P^i$, $\lambda_D^i$ and $\lambda_N^i$), we derive an outer bound for the DoF region of the $K$-user MISO BC. Subsequently, we tighten this outer bound by taking into account a set of inequalities that capture some of the $3^K$ states of the joint CSIT. One of the consequences of this set of inequalities is that for $K\geq3$, it is shown that the DoF region is not completely characterized by the marginal probabilities in contrast to the two-user case. Afterwards, the tightness of these bounds are investigated through the discussion on the achievability. Finally, a two user MIMO BC having CSIT among P and N is considered in which an outer bound for the DoF region is provided and it is shown that in some scenarios it is tight.
\end{abstract}
%Although the results by Tandon et al. show that for the symmetric two user MISO BC (i.e., $\lambda_Q^i=\lambda_Q,\  \forall i\in \{1,2\}, Q\in \{P,D,N\}$), the Degrees of Freedom (DoF) region depends only on the marginal probabilities, we show that this interesting result does not hold in general when $K\geq3$. In other words, the DoF region is a function of all the joint probabilities.
%\begin{IEEEkeywords}
%MISO BC, Alternating CSIT, Degrees of Freedom, Outer Bound, CSIT Pattern
%\end{IEEEkeywords}

\section{Introduction}
In contrast to point to point multiple-input multiple-output (MIMO) communication where the channel state information at the transmitter (CSIT) does not affect the multiplexing gain, in a multiple-input single-output (MISO) broadcast channel (BC), knowledge of CSIT is crucial for interference mitigation and beamforming purposes \cite{Bruno}. However, the assumption of perfect CSIT may not always be true in practice due to channel estimation error and feedback latency. Therefore, the idea of communication under some sort of imperfection in CSIT has gained more attention recently. The so called MAT algorithm was presented in \cite{MAT} where it was shown that in terms of the degrees of freedom, even an outdated CSIT can result in significant performance improvement in comparison to the case with no CSIT.
 Assuming correlation between the feedback information and current channel state (e.g., when the feedback latency is smaller than the coherence time of the channel), the authors in \cite{Gesbert} and \cite{Gou12} consider the degrees of freedom in a time correlated MISO BC which is shown to be a combination of zero forcing beamforming (ZFBF) and MAT algorithm. Following these works, the general case of mixed CSIT and the $K$-user MISO BC with time correlated delayed CSIT are discussed in \cite{Chen12a} and \cite{xinping_Kuser}, respectively. While all these works consider the concept of delayed CSIT in time domain, \cite{Chenxi} and \cite{Hao} deal with the DoF region and its achievable schemes in a frequency correlated MISO BC where there is no delayed CSIT but imperfect CSIT across subbands, which is more inline with practical systems as Long Term Evolution (LTE) \cite{Bruno}.
  In \cite{Tandon}, the synergistic benefits of alternating CSIT over fixed CSIT was presented in a two user MISO BC with two transmit antennas. In \cite{Varanasi} and \cite{Amuru}, the MISO BC with hybrid CSIT (Perfect or Delayed) was considered. The recent work of \cite{Lashgari} investigates the DoF region of the K-user MISO BC with hybrid CSIT and linear encoding at the transmitter. \cite{Lee1} and \cite{Lee2} show that the optimal sum DoF is achievable if the CSIT is not too delayed in broadcast channels and interference networks, respectively.
%  It is important to note that the current paper deals with a more general scenario in the sense that 1) The CSIT could be unknown (N) 2) The CSIT availability could be fixed, alternating or a mixture (for a subset of users fixed and for the remaining subset alternating).
%The converse in \cite{Tandon} is based on the idea of assigning artificial receivers to the users whose observations are (statistically) equivalent to the corresponding user when CSIT is (not) perfect. However, whether this brilliant approach could be generalized to the scenarios with more than two transmit antennas and two users is unknown. Therefore, for such scenarios, it becomes necessary to check other ways to find the fundamental limits of the system.

The complete characterization of the general MISO BC with perfect, delayed or unknown CSIT is an open problem. The main aim of this paper is to investigate this problem and provide some answers toward this goal.
To this end, our contributions are as follows.
\begin{itemize}
  \item Given the marginal probabilities of CSIT in a $K$-user MISO BC, we derive an outer bound for the DoF region.
  \item A set of inequalities is proposed that captures not only the marginals, but also the joint CSIT distribution. This shows that for the K-user case ($K\geq 3$), marginal probabilities are not sufficient for characterizing the DoF region.
  \item  The tightness of the outer bounds is investigated in certain cases.
  \item Finally, a two-user MIMO BC is considered in which the CSI of a user is either perfect or unknown. An outer bound for the DoF region is provided and it is shown to be tight when the joint CSIT probabilities satisfy a certain relationship.
\end{itemize}

The paper is organized as follows. In section \ref{s2} the system model and preliminaries are presented. An outer bound is provided in section \ref{s3} based on the marginal probabilities and the proof is given in section \ref{t1}. Section \ref{sh} provides an outer bound that depends on the joint CSIT probabilities. The tightness of the outerbounds will be discussed in section \ref{s55}.  Section \ref{MIMO} investigates a two user MIMO BC with CSIT either perfect or unknown, and section \ref{s7} concludes the paper.

Throughout the paper, $f\sim o(\log P)$ is equivalent to $\lim_{P\to \infty}\frac{f}{\log P}=0$. $(.)^T$ and $(.)^H$ denote the transpose and conjugate transpose, respectively.  $CN(\textbf{0},\mathbf{\Sigma})$ is the circularly symmetric complex Gaussian distribution with covariance matrix $\mathbf{\Sigma}$. For a pair of integers $m\leq q$, the discrete interval is defined as $[m:q]=\{m,m+1,\ldots,q\}$. $Y_{[i:j]}=\{Y_i,Y_{i+1},\ldots,Y_j\}$, $Y([i:j])=\{Y(i),Y(i+1),\ldots,Y(j)\}$ and $Y^n=Y([1:n])$.
\section{System Model}\label{s2}
We consider a MISO BC, in which a base station with $M$ antennas sends independent messages $W_1,\ldots,W_K$ to $K$ single-antenna users ($M\geq K$). In a flat fading scenario, the discrete-time baseband received signal of user $k$ at channel use (henceforth, time instant) $t$ can be written as
\begin{equation}\label{equ1}
  Y_k(t)=\mathbf{H}_{k}^{H}(t)\mathbf{X}(t) + W_k(t) \ ,\ k\in[1:K]\ ,\ t\in[1:n]
\end{equation}
where $\mathbf{X}(t)\in C^{(M\times1)}$ is the transmitted signal at time instant $t$ satisfying the (per codeword) power constraint $\sum_{t=1}^n\|\mathbf{x}(t)\|^2\leq nP$. $W_k(t)$ and $\mathbf{H}_{k}(t)$ are the additive noise and channel vector of user $k$, respectively, and are also assumed i.i.d. over the time instants and the users.
%Also, let $H(n)={[\textbf{\textit{h}}_1(n),\ldots,\textbf{\textit{h}}_K(n)]}^H$ and $H^n=\{H(1),\ldots,H(n)\}$.
We assume global perfect Channel State Information at Receivers (CSIR).  %i.e., at time instant $n$, all users have perfect knowledge of $H^n$.

The rate tuple $(R_1,R_2,\ldots,R_K)$, in which $R_i = \frac{\log (|W_i|)}{n}$, is achievable if there exists a coding scheme such that the probability of error in decoding $W_i$ at user $i (i\in[1:K])$ can be made arbitrarily small with sufficiently large coding block length. The DoF region is defined as $\{(d_1,\ldots,d_K)|\exists (R_1,R_2,\ldots,R_K)\in C(P) \mbox{\ such that\ } d_i=\lim_{P\to \infty}\frac{R_i}{\log P},\ \forall i\}$ where $C(P)$ is the capacity region (i.e., the closure of the set of achievable rate tuples).
\begin{figure}[t]
  \centering
  \includegraphics[width=8cm]{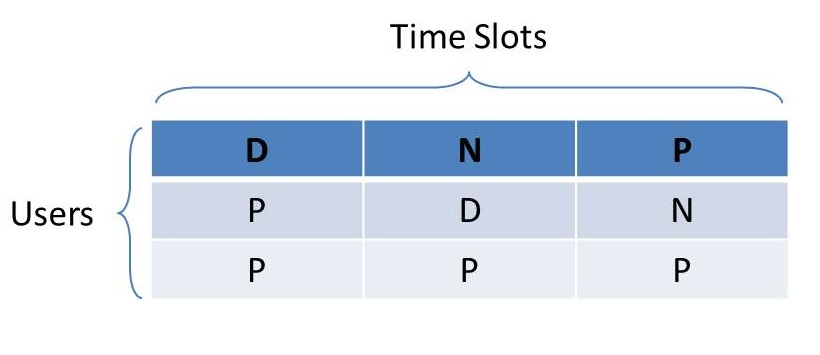}\\
  \caption{A CSIT pattern with $\lambda_{DPP}=\lambda_{NDP}=\lambda_{PNP}=\frac{1}{3}$}\label{fig7}
\end{figure}

The probabilistic model used in this paper for CSIT availability allows the transmitter to have a Perfect (P) instantaneous knowledge of the CSI of a particular user at some time instants, whereas at some other time instants it receives the CSI with Delay (D) and finally, for the remaining time instants the CSI of the user is Not known (N) at the transmitter. The CSIT model can be fixed (i.e., as in the hybrid model), alternating or both (i.e., fixed for a subset of the users and alternating for the remaining subset.) When there is delayed CSIT, we assume that the feedback delay is much larger than the coherence time of the channel making the feedback information completely independent of the current channel state. In this configuration, the joint CSIT of all the $K$ users has at most $3^K$ states. For example, in a 3 user MISO BC, they will be $PPP,PPD,PPN,PDP,\ldots$  with corresponding probabilities $\lambda_{PPP}, \lambda_{PPD},\lambda_{PPN},\lambda_{PDP},\ldots$\ and, as an example, the marginal probability of perfect CSIT for user 1 is $\lambda_P^1=\sum_{Q,Q'\in\{P,D,N\}}\lambda_{PQQ'}$.

By CSIT pattern we refer to the knowledge of CSIT represented in a space-time matrix where the rows and columns represent users and time slots, respectively. The channel remains fixed within each time slot, while it changes independently from one slot to another. For simplicity, we assume the delayed CSI arrives at the transmitter after one time slot. Figure \ref{fig7} shows an example of a CSIT pattern, in which the transmitter knows the channels of users 2 and 3 perfectly at time slot 1 and has no information about the channel of user 1. The CSI of user 1 will be known in the next time slot due to feedback delay and is completely independent of the channel in time slot 2.

\begin{figure}
  \centering
  \includegraphics[width=8cm]{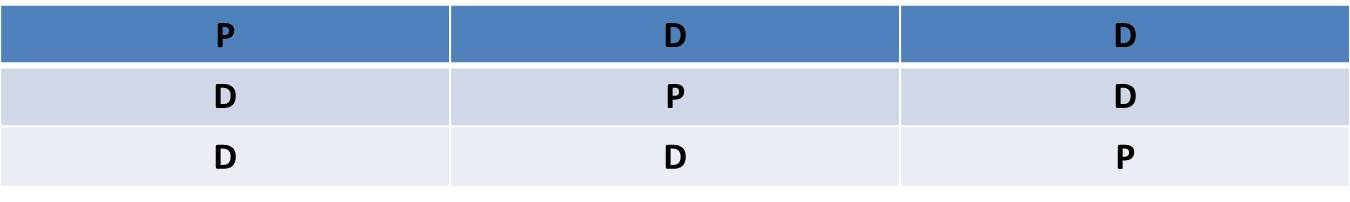}\\
  \caption{A symmetric CSIT pattern for the 3-user MISO BC with the marginals $\lambda_P =\frac{1}{3}, \lambda_D =\frac{2}{3}$.}\label{fig7.}
\end{figure}

Finally, a symmetric CSIT pattern means that the marginal probabilities of perfect, delayed and unknown CSIT are the same across the users, i.e. $\lambda_Q^i=\lambda_Q,\  \forall i\in [1:K], Q\in \{P,D,N\}$. As an example figure \ref{fig7.} shows a symmetric CSIT pattern for the 3-user MISO BC in which $\lambda_P =\frac{1}{3}, \lambda_D =\frac{2}{3}$.
\section{An outer bound given the marginals}\label{s3}
\textbf{Theorem 1}. Let $\pi^j(.)$ be an arbitrary permutation of size $j$ over the indices $(1,2,\ldots,K)$, and $\alpha_{\pi^j}(.)$ be a permutation of $\pi^j$ satisfying\footnote{The reason for arranging the users according to the sum of the perfect and delayed CSIT probabilities becomes clear in (\ref{ee37}).}
\begin{equation}\label{const}
  (\lambda_P^{\alpha_{\pi^j}(i)} + \lambda_D^{\alpha_{\pi^j}(i)})\leq (\lambda_P^{\alpha_{\pi^j}(i+1)} + \lambda_D^{\alpha_{\pi^j}(i+1)})\ \ ,\ \ i\in[1:j-1].
\end{equation}
Given the marginal probabilities of CSIT for user $i$ (which can be any two of $\lambda_P^i, \lambda_D^i$ and $\lambda_N^i$, since $\lambda_P^i + \lambda_D^i + \lambda_N^i = 1$), an outer bound for the DoF region of the $K$-user MISO BC with $M$ transmit antennas at the transmitter ($M\geq K$) is defined by the following sets of inequalities
\begin{align}
 \sum_{i=1}^j\frac{d_{\pi^j(i)}}{i}&\leq 1 + \sum_{i=2}^j\frac{\sum_{r=1}^{i-1}\lambda_P^{\pi^j(r)}}{i(i-1)}\label{theorem1}\\
 \sum_{i=1}^j d_{\pi^j(i)} &\leq 1 + \sum_{i=1}^{j-1}(\lambda_P^{\alpha_{\pi^j}(i)} + \lambda_D^{\alpha_{\pi^j}(i)})\ \ ,\ \ \forall \pi^j,\ j\in[1:K].\label{theorem11}
\end{align}
For the symmetric scenario, the sets of inequalities are simplified as
\begin{align}
\sum_{i=1}^j\frac{d_{\pi^j(i)}}{i}&\leq 1 + \lambda_P\sum_{i=2}^j\frac{1}{i}\\
\sum_{i=1}^j d_{\pi^j(i)} &\leq 1 + (j - 1)(\lambda_P + \lambda_D)\ \ ,\ \ \forall \pi^j,\ j\in[1:K].\label{sym}
\end{align}
For $K=2$, the outer bound boils down to the optimal DoF region in \cite{Tandon}.
\section{Proof of theorem 1}\label{t1}
%The structure of the proof could be briefly itemized as follows.
%\begin{itemize}
%  \item Applying some sort of improvement to the channel.
%  \item The usage of Fano's inequality.
%  \item Application of the \textit{Csisz\'{a}r sum identity} \cite{network_info} as in \cite{Hao} to change the difference between vector entropies into the sum of the component-wise entropy differences.
%  \item Finding an upper bound for these entropy differences by application of two provided lemmas.
%\end{itemize}
%Having an outer bound for the DoF region of the general $K$-user BC ($M\geq K$),
For simplicity, we assume $j = K$, since it is obvious that each subset of users with cardinality $j$ ($j<K$) can be regarded as a $j$-user BC.
%Therefore, we only consider the proof of the inequalities involving all the $K$ users. For simplicity, we show the inequalities for the
Also, we assume the identity permutation (i.e., $\pi^K(i)=i$) while the results could be easily applied to any other arbitrary permutation.
\subsection{Proof of $\sum_{i=1}^K\frac{d_i}{i}\leq 1 + \sum_{i=2}^K\frac{\sum_{r=1}^{i-1}\lambda_P^r}{i(i-1)}$}
First, we improve the channel by giving the message and observation of user $i$ to users $[i+1:K]$ ($i\in[1:K-1]$). Hence, from Fano's inequality,
\begin{equation}\label{Fano}
  nR_i \leq I(W_i;Y_{[1:i]}^n|W_{[1:i-1]},\Omega^n)+n\epsilon_n
\end{equation}
where $\Omega^n$ denotes the global CSIR up to time instant $n$, $W_0=\emptyset$ and $\epsilon_n$ goes to zero as $n$ goes to infinity.
%This improvement does not decrease the capacity region, meaning that the capacity region of the original channel is a subset of this improved channel. Also,
By this improvement, channel input and outputs (i.e., the enhanced observations of users) form a Markov chain which results in a physically degraded broadcast channel \cite{Cover}. Therefore, according to \cite{Elgamal}, since feedback does not increase the capacity of physically degraded broadcast channels, we can ignore the delayed CSIT (D) and replace them with No CSIT (N).
%In other words, at time instant $n$, knowledge of the CSI up to time instant $n-1$ is not beneficial in a physically degraded BC.
This is equivalent to having the channel of user $i$ perfectly known with probability $\lambda_P^i$ and not known otherwise.
%It is important to note that although the channel has become physically degraded, the perfect CSIT (P) cannot be replaced with No CSIT (N), since (P) means that at time instant $n$ the current state of the channel is known to the transmitter perfectly which enables it to know the received signal within noise level (i.e., the results of \cite{Elgamal} cannot be applied in this case.)
From now on, we ignore the term $n\epsilon_n$ for simplicity (since later it will be divided by $n$ and $n\to\infty$) and write
\begin{align}
\sum_{i=1}^K\frac{nR_i}{i}  &\leq  \sum_{i=1}^K\frac{I(W_i;Y_{[1:i]}^n|W_{[1:i-1]},\Omega^n)}{i}  \\
 &\leq h(Y_1^n|\Omega^n) +\sum_{i=2}^K\left[\frac{h(Y_{[1:i]}^n|W_{[1:i-1]},\Omega^n)}{i}\right.\nonumber\\&\ \ \ \left.-\frac{h(Y_{[1:i-1]}^n|W_{[1:i-1]},\Omega^n)}{i-1}\right]+ no(\log P)\label{e1}
\end{align}
where $Y_0=\emptyset$ and we have used the fact that $\frac{h(Y_{[1:K]}^n|W_{[1:K]},\Omega^n)}{nK}\sim o(\log P)$, since with the knowledge of $W_{[1:K]}$ and $\Omega^n$, the observations $Y_{[1:K]}^n$ can be reconstructed within the noise distortion.
%From the chain rule of entropies, each of the terms in the summation in (\ref{e1}) can be written as
%\begin{equation}\label{eq}
%\sum_{t=1}^n\left[\frac{h(Y_{[1:i]}(t)|W_{[1:i-1]},Y_{[1:i]}^{t-1},\Omega^t)}{i}-\frac{h(Y_{[1:i-1]}(t)|W_{[1:i-1]},Y_{[1:i-1]}^{t-1},\Omega^t)}{i-1}\right].
%\end{equation}
%%where $Y_i^{t-1}$ is the time extension of $Y$ from time instant $i$ to $t-1$.
%By adding $Y_i^{t-1}$ to the conditions of the second entropy, (\ref{eq})  will be increased. Therefore,
%%\begin{equation}
%%\sum_{t=1}^n\left[\frac{h(Y_{[1:i]}(t)|T_{i,t},\Omega(t))}{i}-\frac{h(Y_{[1:i-1]}(t)|T_{i,t},\Omega(t))}{i-1}\right]
%%\end{equation}
%\begin{equation}\label{e30}
%    \sum_{i=1}^K\frac{nR_i}{i}\leq \underbrace{h(Y_1^n|\Omega^n)}_{\leq n\log P}+\sum_{i=2}^K\sum_{t=1}^n\left[\frac{h(Y_{[1:i]}(t)|U_{i,t},\Omega(t))}{i}-\frac{h(Y_{[1:i-1]}(t)|U_{i,t},\Omega(t))}{i-1}\right]+no(\log P)
%\end{equation}
%where $U_{i,t}=(W_{[1:i-1]},Y_{[1:i]}^{t-1},\Omega^{t-1})$ and $\Omega(t)$ is the global CSIR at time instant $t$.
Before going further, the following lemma is needed.

\textbf{Lemma 1}. Let $\Gamma_N=\{Y_1,Y_2,\ldots,Y_N\}$ be a set of $N(\geq2)$ arbitrary random variables and $\Psi_i^{j}(\Gamma_N)$ be a sliding window of size $j$ over $\Gamma_N$ ($1\leq i,j \leq N$) starting from $Y_i$ i.e.,
\[\Psi_i^{j}(\Gamma_N) = Y_{(i-1)_N+1},Y_{(i)_N+1},\ldots,Y_{(i+j-2)_N+1}\]
where $(.)_N$ defines the modulo $N$ operation. Then, the following inequality holds for $\forall m\in[1:N-1]$
\begin{equation}\label{e..6}
  (N-m)h(Y_{[1:N]}|A)\leq \sum_{i=1}^{N}h(\Psi_i^{N-m}(\Gamma_N)|A)
\end{equation}
where $A$ is an arbitrary condition.
\begin{proof} We prove the lemma by showing that for every fixed $m(\geq 1)$, (\ref{e..6}) holds for all $N(\geq m+1)$ using induction. It is obvious that for every $m(\geq 1)$, (\ref{e..6}) holds for $N=m+1$. In other words, $h(Y_{[1:N]}|A)\leq \sum_{i=1}^{N}h(Y_i|A)$. Now, considering that (\ref{e..6}) is valid for $N(\geq m+1)$, we show that it also holds for $N+1$. Replacing $N$ with $N+1$, we have
\begin{align}
&(N+1-m)h(Y_{[1:N+1]}|A)\nonumber\\&=h(Y_{[1:N+1]}|A)+\!(N-m)h(Y_{[1:N-1]},\overbrace{Y_N,Y_{N+1}}^{Z}|A)\nonumber\\
&\leq h(Y_{[1:N+1]}|A)+\sum_{i=1}^{N}h(\Psi_i^{N-m}(\Phi_N)|A)\label{e..8}\\
&= h(Y_{[1:N+1]}|A)+\sum_{i=1}^{m}h(\Psi_i^{N-m}(\Phi_N)|A)\nonumber\\&\ \ \ + \sum_{i=m+1}^{N}h(\Psi_i^{N+1-m}(\Gamma_{N+1})|A)\label{e..9}\\
&= h(Y_{[N-m+1:N]}|Y_{N+1},Y_{[1:N-m]},A)\nonumber\\&\ \ \ +\sum_{i=1}^{m}h(\Psi_i^{N-m}(\Phi_N)|A)+h(Y_{N+1},Y_{[1:N-m]}|A)\nonumber\\&\ \ \ +\sum_{i=m+1}^{N}h(\Psi_i^{N+1-m}(\Gamma_{N+1})|A)\label{e..11}\\
&= h(Y_{[N-m+1:N]}|Y_{N+1},Y_{[1:N-m]},A)\nonumber\\&\ \ \ +\sum_{i=1}^{m}h(\Psi_i^{N-m}(\Phi_N)|A)+\sum_{i=m+1}^{N+1}h(\Psi_i^{N+1-m}(\Gamma_{N+1})|A)\nonumber\\
&= \sum_{i=1}^mh(Y_{N-m+i}|Y_{N+1},Y_{[1:N-m+i-1]},A)\nonumber\\&\ \ \ +\sum_{i=1}^{m}h(Y_{[i:N-m+i-1]}|A)+\sum_{i=m+1}^{N+1}h(\Psi_i^{N+1-m}(\Gamma_{N+1})|A)\label{e..13}\\
&\leq \sum_{i=1}^mh(Y_{N-m+i}|Y_{[i:N-m+i-1]},A)+\sum_{i=1}^{m}h(Y_{[i:N-m+i-1]}|A)\nonumber\\&\ \ \ +\sum_{i=m+1}^{N+1}h(\Psi_i^{N+1-m}(\Gamma_{N+1})|A)\label{e13..75}\\
&= \sum_{i=1}^{m}h(\Psi_i^{N+1-m}(\Gamma_{N+1})|A)+\sum_{i=m+1}^{N+1}h(\Psi_i^{N+1-m}(\Gamma_{N+1})|A) \nonumber\\
&= \sum_{i=1}^{N+1}h(\Psi_i^{N+1-m}(\Gamma_{N+1})|A)
\end{align}
%\begin{align}
%  &= \sum_{i=1}^{m}h(\Psi_i^{N+1-m}(\Gamma_{N+1})|A)+\sum_{i=m+1}^{N+1}h(\Psi_i^{N+1-m}(\Gamma_{N+1})|A) \label{e14}\\
%  &= \sum_{i=1}^{N+1}h(\Psi_i^{N+1-m}(\Gamma_{N+1})|A)
%\end{align}
where in (\ref{e..8}), $\Phi_N=\{Y_{[1:N-1]},Z\}$ and we have used the validity of (\ref{e..6}) for $N$. In (\ref{e..9}), we have used the fact that $\Psi_i^{N+1-m}(\Gamma_{N+1})=\Psi_i^{N-m}(\Phi_N)$ for $i \in [m+1:N]$ . In (\ref{e..11}), the chain rule of entropies is used and in (\ref{e..13}), the sliding window is written in terms of its elements. Finally, in (\ref{e13..75}), the fact that conditioning reduces the differential entropy is used. Therefore, since $m(\geq 1)$ was chosen arbitrarily and (\ref{e..6}) is valid for $N=m+1$ and from its validity for $N(\geq m+1)$ we could show it also holds for $N+1$, we conclude that (\ref{e..6}) holds for all values of $m$ and $N$ satisfying $1\leq m\leq N-1$. \qedhere
\end{proof}
Each term in the summation of (\ref{e1}) can be rewritten as
\begin{equation*}
  \frac{(i-1)h(Y_{[1:i]}^n|W_{[1:i-1]},\Omega^n)-ih(Y_{[1:i-1]}^n|W_{[1:i-1]},\Omega^n)}{i(i-1)}
\end{equation*}
\begin{align}
  &\leq \frac{\sum_{r=1}^i\left[h(\Psi_r^{i-1}(\Gamma_i)|T_{i,n})-h(Y_{[1:i-1]}^n|T_{i,n})\right]}{i(i-1)}\label{e.3e} \\
  &= \frac{\sum_{r=1}^{i-1}\left[h(Y_i^n|E_{r,i},T_{i,n})-h(Y_r^n|E_{r,i},T_{i,n})\right]}{i(i-1)}\label{e3e}
\end{align}
where $\Gamma_i=\{Y_{[1:i]}^n\}$, $T_{i,n}=\{W_{[1:i-1]},\Omega^n\}$ and $E_{r,i} = \{Y_{[1:i-1]}^n\}-\{Y_r^n\}$. (\ref{e.3e}) is from the application of lemma 1 ($m=1$) and (\ref{e3e}) is from the chain rule of entropies. Before going further, the following lemma is needed. This lemma, which is based on \cite{Jafar}, is the key part in the proof.

\textbf{Lemma 2.} In the $K$-user MISO BC defined in (\ref{equ1}), for the users $m,q\in[1:K]$ ($m\neq q$), we have
%\begin{eqnarray}
%% \nonumber to remove numbering (before each equation)
%  Y_m(j) &=& \textit{\textbf{h}}_m(j)^T\textit{\textbf{x}}(j)+w_m(j) \\
%  Y_q(j) &=& \textit{\textbf{h}}_q(j)^T\textit{\textbf{x}}(j)+w_q(j).
%\end{eqnarray}
%Without loss of generality, we assume $m > q$. For simplicity, we assume that the communication is done in real dimensions where $\textit{\textbf{x}}\in R^{M\times 1}$ satisfying $E\left[\|\textit{\textbf{x}}\|^2\right]\leq P$, $\textit{\textbf{h}}_m$ and $\textit{\textbf{h}}_q$ have the distribution $N(\textbf{0},\textbf{I})$ and $w_m$ and $w_q$ have the distribution $N(0,1)$. When the CSIT of a user is either Perfect (P) or Not known (N), the following upper bound holds for the difference between entropies
\begin{equation}\label{e10}
 \lim_{n,P\to \infty} \frac{h(Y_m^n|A)-h(Y_q^n|A)}{n\log P}\leq \left\{\begin{array}{cc} 1 & \mbox{CSIT of } q \mbox{ is } P \\ 0 & \mbox{CSIT of } q \mbox{ is } N  \end{array}\right.
\end{equation}
where $A$ is a condition such as the condition of entropies in (\ref{e3e}) or later in (\ref{eq1}). Interestingly, (\ref{e10}) is only a function of the CSIT of the second user.
%In other words, in the four possible cases of $PP,PN,NP$ and $NN$, the upper bound (not the exact value) for the pre-log factor of the difference is defined by the CSIT of the second user resulting in the same upper bound for the $PN$ or $NN$ case, and the same upper bound for the $PP$ or $NP$ case.
\begin{proof}
Based on the four possible states for the joint CSIT of $m$ and $q$, we have
\subsubsection{CSIT of m is N or P and CSIT of q is P}
\begin{equation}\label{e14}
  h(Y_m^n|A)-h(Y_q^n|A) \leq \underbrace{h(Y_m^n|A)}_{\leq n\log (P)}-\underbrace{h(Y_q^n|A,W_{[1:K]})}_{no(\log P)}
\end{equation}
 A Gaussian input with the conditional covariance matrix of $\Sigma_{X|A}=P\textbf{{u}}_q^{\perp}{\textbf{{u}}_q^{\perp}}^H$ achieves the upper bound, where $\textbf{{u}}_q^{\perp}$ is a unit vector in the direction orthogonal to $\textbf{{H}}_q$ (since $\textbf{{H}}_q$ is known).
\subsubsection{CSIT of m is N and CSIT of q is N}
In this case both $Y_m^n$ and $Y_q^n$ are statistically equivalent (i.e., having the same probability density functions, and subsequently, the same entropies.) Therefore,
\begin{equation}
  h(Y_m^n|A)-h(Y_q^n|A)=0
\end{equation}
\subsubsection{CSIT of m is P and CSIT of q is N}
This is the second result of Theorem 1 in \cite{Jafar}\footnote{The differential entropy terms in the left hand side of (\ref{e10}) can be written in terms of the
expectation of the difference of entropies conditioned on the realizations of A. Since the conditional
probability density functions exist and have a bounded peak, the same steps of \cite{Jafar} as discretization,
considering the cannonical form and bounding the cardinality of aligned image set can be applied.}.
\end{proof}
From (\ref{e1}) and (\ref{e3e}), we have
\begin{align}
  \sum_{i=1}^K\! \frac{nR_i}{i} &\leq\! \sum_{i=2}^K\sum_{r=1}^{i-1}\frac{h(Y_i^n|A_{r,i})-h(Y_r^n|A_{r,i})}{i(i-1)}\nonumber\\&\ \ \ +n\log P +no(\log P)\nonumber\\
%&= n\log P + \sum_{i=2}^K\sum_{r=1}^{i-1}\frac{\sum_{t=1}^n\left[h(Y_i(t)|A(r,i,t))-h(Y_r(t)|A(r,i,t))\right]}{i(i-1)}+no(\log P)\label{e47e} \\
&\leq n\log P + \sum_{i=2}^K\sum_{r=1}^{i-1}\frac{n\lambda_P^r}{i(i-1)}\log P +no(\log P) \label{e4e}
\end{align}
where $A_{r,i}$ is the condition of the entropies in (\ref{e3e}) and (\ref{e4e}) is from the application of lemma 2 and the fact that $n$ is sufficiently large. Therefore,
\begin{equation}
  \sum_{i=1}^K\frac{d_i}{i}\leq 1 + \sum_{i=2}^K\frac{\sum_{r=1}^{i-1}\lambda_P^r}{i(i-1)}.
\end{equation}
It is obvious that the same approach can be applied to any other permutations on $(1,2,\ldots,K)$ which results in (\ref{theorem1}).
In addition to the mentioned proof, an alternative proof is provided in Appendix \ref{s4}.
\subsection{Proof of $\sum_{i=1}^K d_i \leq 1 + \sum_{i=1}^{K-1}(\lambda_P^{\alpha_{\pi^K}(i)} + \lambda_D^{\alpha_{\pi^K}(i)})$}\label{enh}
We enhance the channel in two ways:
\begin{enumerate}
  \item Like the approach in \cite{Tandon}, whenever there is delayed CSIT ($D$), we assume that it is perfect instantaneous CSIT ($P$), but we keep the probability of delayed CSIT. In other words, the CSIT of user $i$ is perfect with probability $\lambda_P^i+\lambda_D^i$ and unknown otherwise.
  \item We give the message of user $i$ to users $[i+1:K]$.
\end{enumerate}
  Therefore,
  \begin{equation}\label{e3}
    nR_i\leq I(W_i;Y_i^n|W_{[1:i-1]},\Omega^n)+n\epsilon_n\ ,\ \forall i\in[1:K].
  \end{equation}
   By summing (\ref{e3}) over users and writing the mutual information in terms of differential entropies,
   \begin{align}
     \sum_{i=1}^KnR_i&\leq \overbrace{h(Y_1^n|\Omega^n)}^{\leq n\log P}+ no(\log P)\nonumber\\&\ \ \ + \sum_{i=2}^K\left[h(Y_i^n|W_{[1:i-1]},\Omega^n)-h(Y_{i-1}^n|W_{[1:i-1]},\Omega^n)\right] \label{eq1}.
  \end{align}
%The term in the summation could be written as
%  \begin{equation}\label{e33}
%   \sum_{t=1}^n\left[h(Y_i(t)|F_{i,t},\Omega(t))-h(Y_{i-1}(t)|F_{i,t},\Omega(t))\right]
%  \end{equation}
%where $F_{i,t}=\left (W_{[1:i-1]},\Omega^{t-1},Y_{i-1}^{t-1},Y_i([t+1:n])\right)$.
%The proof of (\ref{e33}) is provided in Appendix \ref{s5}.
%Therefore,
% \begin{align}
%   \sum_{i=1}^KnR_i&\leq n\log P +\sum_{i=2}^K\sum_{t=1}^n\left[h(Y_i(t)|F_{i,t},\Omega(t))-h(Y_{i-1}(t)|F_{i,t},\Omega(t))\right]+ no(\log P)\label{eq1}
% \end{align}
%and finally,
By applying the results of lemma 2 to (\ref{eq1}), we have
\begin{equation}\label{e373}
   \sum_{i=1}^K d_i \leq 1+\sum_{i=2}^K(\lambda_P^{i-1}+\lambda_D^{i-1})=1+\sum_{i=1}^{K-1}(\lambda_P^{i}+\lambda_D^{i}).
\end{equation}
Let $\pi^K(.)$ be an arbitrary permutation of size $K$ on $(1,\ldots,K)$. Applying the same reasoning, we have
\begin{equation}\label{e37}
   \sum_{i=1}^K d_i \leq 1+\sum_{i=1}^{K-1}(\lambda_P^{\pi^K(i)}+\lambda_D^{\pi^K(i)})\ \ ,\ \ \forall \pi^K(.).
\end{equation}
(\ref{e37}) results in $K$ inequalities all having the same left hand side. Therefore,
%\begin{equation}\label{e37}
%   \sum_{i=1}^K d_i \leq 1+\sum_{i=1}^{K-1}(\lambda_P^{\pi^K(i)}+\lambda_D^{\pi^K(i)})\ \ ,\forall \pi^K(.)
%\end{equation}
%(\ref{e37}) results in $K$ inequalities all having the same left hand side. Therefore,
\begin{equation}\label{ee37}
   \sum_{i=1}^K d_i \leq 1+\min_{\pi^K(.)}{\sum_{i=1}^{K-1}(\lambda_P^{\pi^K(i)}+\lambda_D^{\pi^K(i)})}
\end{equation}
This is due to the possible orders of channel enhancements and it is obvious that $\alpha_{\pi^K}(.)$ will minimize (\ref{ee37}) if it satisfies (\ref{const}) (for $j=K$.)
\section{An outer bound capturing the joint CSIT probabilities}\label{sh}

%Here, we show that two different CSIT patterns, though having the same marginal probabilities, do not necessarily have the same DoF region. Consider the two simple symmetric CSIT patterns shown in figure \ref{fig100}. According to the theorem, the DoF region of both has an outer bound with the corner points $(1,\frac{1}{3},\frac{1}{3}),(\frac{1}{3},1,\frac{1}{3})$ and $(\frac{1}{3},\frac{1}{3},1)$. It is obvious that the corner points are achievable for pattern $(a)$, and in what follows we show that they are not achievable for pattern $(b)$.

\begin{figure}[t]
  \centering
  \includegraphics[width=8cm]{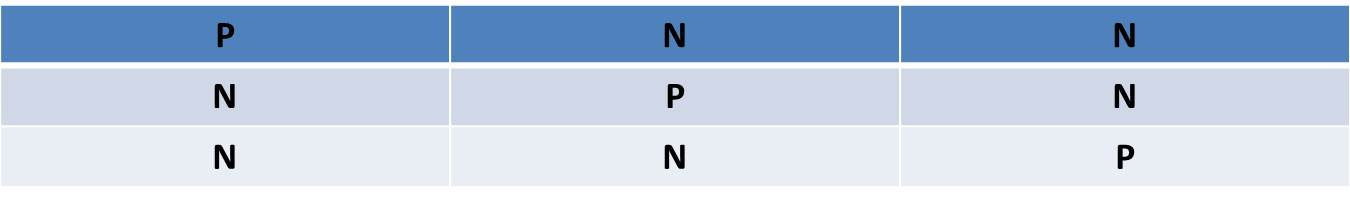}\\
  \caption{A symmetric CSIT pattern for the 3-user MISO BC.}\label{fig100}
\end{figure}

In the previous section, an outer bound was provided in terms of the marginal probabilities. In this section, we tighten the outer bound by introducing a set of inequalities that captures the joint CSIT probabilities. We start with simple motivating examples. Consider the pattern shown in figure \ref{fig100}. By Fano's inequality, we write,
\begin{align}
  nR_1 &\leq I(W_1;Y_1^n|\Omega^n) \label{eq14}\\
  nR_1 &\leq I(W_1;Y_1^n|\Omega^n,W_2).\label{eq15}
\end{align}
Adding (\ref{eq14}) and (\ref{eq15}) results in
\begin{equation}\label{eq16}
  2nR_1 \leq I(W_1;Y_1^n|\Omega^n)+ I(W_1;Y_1^n|\Omega^n,W_2).
\end{equation}
By doing the same for $R_2$, we have
\begin{equation}\label{eq17}
  2nR_2 \leq I(W_2;Y_2^n|\Omega^n)+ I(W_2;Y_2^n|\Omega^n,W_1).
\end{equation}
Finally, the rate of user 3 is written as
\begin{equation}\label{eq18}
  nR_3 \leq I(W_3;Y_3^n|\Omega^n,W_1,W_2).
\end{equation}
Therefore,
\begin{align}
  &2nR_1+2nR_2+nR_3\nonumber\\
  &\leq \underbrace{h(Y_2^n|\Omega^n,W_1)-h(Y_1^n|\Omega^n,W_1)}_{\leq \frac{n}{3}\log P}+ h(Y_3^n|\Omega^n,W_1,W_2) \nonumber \\
  &\ \ \underbrace{+h(Y_1^n|\Omega^n,W_2)-h(Y_2^n|\Omega^n,W_2)}_{\leq \frac{n}{3}\log P}+\underbrace{h(Y_1^n|\Omega^n)}_{\leq n\log P}+\underbrace{h(Y_2^n|\Omega^n)}_{\leq n\log P}\nonumber\\
  &\ \ \underbrace{-h(Y_1^n|\Omega^n,W_1,W_2)-h(Y_2^n|\Omega^n,W_1,W_2)}_{\leq -h(Y_1^n,Y_2^n|\Omega^n,W_1,W_2)} \label{53}\\
  &\leq \frac{8n}{3}\log P + h(Y_3^n|\Omega^n,W_1,W_2)-h(Y_1^n,Y_2^n|\Omega^n,W_1,W_2)\label{mohem}\\
  &= \frac{8n}{3}\log P+\underbrace{h(Y_3^n|\Theta)-h(Y_{2,PNN}^n,Y_{1,NPN}^n,Y_{1,NNP}^n|\Theta)}_{o(\log P)}\nonumber\\
  &\ \underbrace{-h(Y_{1,PNN}^n,Y_{2,NPN}^n,Y_{2,NNP}^n|\Theta,Y_{2,PNN}^n,Y_{1,NPN}^n,Y_{1,NNP}^n)}_{\leq -h(Y_{1,PNN}^n,Y_{2,NPN}^n,Y_{2,NNP}^n|\Theta,Y_{2,PNN}^n,Y_{1,NPN}^n,Y_{1,NNP}^n,W_3)\sim o(\log\!P)}\label{a2}\\
  &\leq \frac{8n}{3}\log P
\end{align}
where in (\ref{53}), lemma 2 is applied to the differences resulting in the values written under the braces and in (\ref{a2}), $\Theta=\{\Omega^n,W_1,W_2\}$. We have split the observations of users 1 and 2 in terms of the joint CSIT, i.e., $Y_1^n=(Y_{1,PNN}^n,Y_{1,NPN}^n,Y_{1,NNP}^n)$ and $Y_2^n=(Y_{2,PNN}^n,Y_{2,NPN}^n,Y_{2,NNP}^n)$. (\ref{a2}) is due to the fact that there is at least one unknown CSIT (N) in the joint states of user 1 and user 2 (i.e., PN, NP and NN. see rows 1 and 2 of the CSIT pattern shown in figure \ref{fig100}). Therefore, we have the following inequalities for the pattern shown in figure \ref{fig100}
\begin{align}
  2d_1+2d_2+d_3 &\leq \frac{8}{3} \nonumber\\
  2d_1+d_2+2d_3 &\leq \frac{8}{3} \nonumber\\
  d_1+2d_2+2d_3 &\leq \frac{8}{3}.\label{ew1}
\end{align}
From (\ref{ew1}), the sum DoF of the pattern in figure \ref{fig100} has the upper bound of $\frac{8}{5}$, while it can be easily verified that for the pattern with PPP in the first slot and NNN in the next two slots, which has the same marginals as in figure \ref{fig100}, the sum DoF is $\frac{5}{3}(> \frac{8}{5})$. This simple example confirms that for the K-user MISO BC ($K\geq 3$), the marginal probabilities are not sufficient in characterizing the DoF region\footnote{It is important to emphasize on the difference between the following two statements

a)	Two CSIT patterns with \textbf{different marginals} can have the \textbf{same} DoF regions.

b)	Two CSIT patterns with \textbf{the same marginals} can have \textbf{different} DoF regions.

 The first statement is already known in literature. For example, by comparing the original 2-user MAT (i.e., $\lambda_D=1$) and the scheme DN,ND,NN in [9], it is concluded that both of them have the sum DoF of 4/3, while having different marginal prbabilities (for the latter, $\lambda_D=\frac{1}{3}$). However, the set of inequalities proposed in this section addresses the second statement which is a new problem and cannot result from the first statement.}.
Motivated by this simple example, we can have the following set of inequalities for the 3-user MISO BC with P and N
%\begin{align}
%  2nR_1+2nR_2+nR_3 &\leq \underbrace{h(Y_1^n|H^n)}_{\leq n\log P}+\underbrace{h(Y_2^n|H^n)}_{\leq n\log P}  \\
%  &\ \underbrace{+h(Y_2^n|H^n,W_1)-h(Y_1^n|H^n,W_1)}_{\leq n(\lambda_P^1+\lambda_D^1)\log P}\underbrace{+h(Y_1^n|H^n,W_2)-h(Y_2^n|H^n,W_2)}_{\leq n(\lambda_P^2+\lambda_D^2)\log P}\\
%  &\ \underbrace{+ h(Y_3^n|H^n,W_1,W_2)\underbrace{-h(Y_1^n|H^n,W_1,W_2)-h(Y_2^n|H^n,W_1,W_2)}_{\leq -h(Y_1^n,Y_2^n|H^n,W_1,W_2)}}_{\leq n(\lambda_{PP-}+\lambda_{PD-}+\lambda_{DP-}+\lambda_{DD-})\log P}
%\end{align}
\begin{align}\label{ew2}
  2d_1+2d_2+d_3 &\leq 2+2\lambda_P+\lambda_{PP-}\nonumber \\
  2d_1+d_2+2d_3 &\leq 2+2\lambda_P+\lambda_{P-P}\nonumber\\
  d_1+2d_2+2d_3 &\leq 2+2\lambda_P+\lambda_{-PP}
\end{align}
where a dashed line in the above means that the CSIT of the corresponding user is not important (for example, $\lambda_{PP-}=\lambda_{PPP}+\lambda_{PPN}$ which is a summation over all the possible states for the CSIT of user 3).
By looking at the difference of entropies in (\ref{mohem}), it is observed that this difference is of order $o(\log P)$ when there is at least one N in the joint CSIT of users 1 and 2 (i.e., PNN, PNP, NPN, NPP, NNP and NNN) and, therefore, is upperbounded by $n(\lambda_{PPP}+\lambda_{PPN})\log P$. This results in the first inequality of (\ref{ew2}) and the same reasoning applies to the remaining two inequalities. (\ref{ew2}) is a set of inequalities that captures the joint CSIT probabilities and is not only a function of the marginals.

\begin{figure}
  \centering
  \includegraphics[width=8cm]{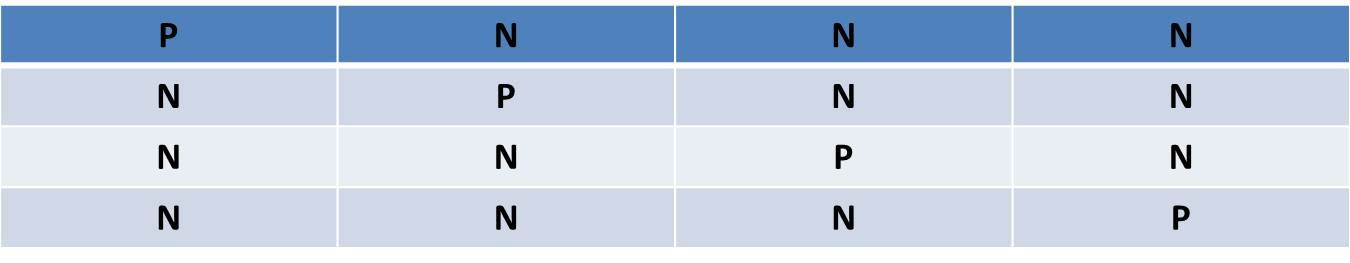}\\
  \caption{A symmetric CSIT pattern for the 4-user MISO BC.}\label{fig101}
\end{figure}

Now consider the pattern shown in figure \ref{fig101} for the 4-user MISO BC.
%We can write
%\begin{align}
%  nR_1 &\leq I(W_1;Y_1^n|\Omega^n) \label{e1}\\
%  nR_1 &\leq I(W_1;Y_1^n|\Omega^n,W_2)\label{e2}.
%\end{align}
%Adding (\ref{e1}) and (\ref{e2}) results in
%\begin{equation*}
%  2nR_1 \leq I(W_1;Y_1^n|\Omega^n)+ I(W_1;Y_1^n|\Omega^n,W_2).
%\end{equation*}
%By doing the same for $R_2$, we have
%\begin{equation*}
%  2nR_2 \leq I(W_2;Y_2^n|\Omega^n)+ I(W_2;Y_2^n|\Omega^n,W_1).
%\end{equation*}
%For user 3, we can write
%\begin{equation*}
%  2nR_3 \leq I(W_3;Y_3^n|\Omega^n,W_1,W_2)+ I(W_3;Y_3^n|\Omega^n,W_1,W_2).
%\end{equation*}
From (\ref{eq16}), (\ref{eq17}) and (\ref{eq18}), we can write
\begin{align}
  &2n(R_1+R_2+R_3)\nonumber\\&\leq \underbrace{h(Y_2^n|\Omega^n,W_1)-h(Y_1^n|\Omega^n,W_1)}_{\leq \frac{n}{4}\log P} \nonumber \\
  &\ \ \underbrace{+h(Y_1^n|\Omega^n,W_2)-h(Y_2^n|\Omega^n,W_2)}_{\leq \frac{n}{4}\log P}+\underbrace{h(Y_1^n|\Omega^n)}_{\leq n\log P}+\underbrace{h(Y_2^n|\Omega^n)}_{\leq n\log P}\nonumber\\
  &\ \ \underbrace{+h(Y_3^n|\Omega^n,W_1,W_2)-h(Y_1^n|\Omega^n,W_1,W_2)}_{\leq \frac{n}{4}\log P} \nonumber\\&\ \ \ \underbrace{+h(Y_3^n|\Omega^n,W_1,W_2)-h(Y_2^n|\Omega^n,W_1,W_2)}_{\leq \frac{n}{4}\log P}\nonumber\\
  &\ \ \ -2h(Y_3^n|\Omega^n,W_1,W_2,W_3)\nonumber\\
  &\leq 3n\log P -2h(Y_3^n|\Omega^n,W_1,W_2,W_3)\label{e3}
\end{align}
Alternatively, we can change the role of users 1 and 3 and write
\begin{align*}
    2nR_1 &\leq I(W_1;Y_1^n|\Omega^n,W_2,W_3)+ I(W_1;Y_1^n|\Omega^n,W_2,W_3)\\
    2nR_2 &\leq I(W_2;Y_2^n|\Omega^n)+ I(W_2;Y_2^n|\Omega^n,W_3)\\
    2nR_3 &\leq I(W_3;Y_3^n|\Omega^n)+ I(W_3;Y_3^n|\Omega^n,W_2).
\end{align*}
Following the same reasoning in (\ref{e3}), we have
\begin{equation}\label{e4}
    2n(R_1+R_2+R_3)\leq 3n\log P -2h(Y_1^n|\Omega^n,W_1,W_2,W_3).
\end{equation}
Adding (\ref{e3}) and (\ref{e4}), we have
\begin{align}
    &4n(R_1+R_2+R_3)\nonumber\\&\leq 6n\log P \nonumber\\&\ \ \ -2\left(h(Y_1^n|\Omega^n,W_1,W_2,W_3)+h(Y_3^n|\Omega^n,W_1,W_2,W_3)\right)\nonumber\\
    &\leq 6n\log P -2h(Y_1^n,Y_3^n|\Omega^n,W_1,W_2,W_3)\label{e5}.
\end{align}
For the rate of user 4, we can write
\begin{align}
    2nR_4&\leq 2I(W_4;Y_4^n|\Omega^n,W_1,W_2,W_3)\nonumber\\
    &=2h(Y_4^n|\Omega^n,W_1,W_2,W_3)\nonumber\\&\ \ \ \underbrace{-2h(Y_4^n|\Omega^n,W_1,W_2,W_3,W_4)}_{o(\log P)}.\label{e6}
\end{align}
Adding (\ref{e5}) and (\ref{e6}), we get
\begin{align}
    &4n(R_1+R_2+R_3)+2nR_4\nonumber\\&\leq 6n\log P +2\left(h(Y_4^n|\Psi)-h(Y_1^n,Y_3^n|\Psi)\right)\label{ps}\\
    &\leq 6n\log P +\underbrace{2\left(h(Y_4^n|\Psi)-h(T_n|\Psi)\right)}_{o(\log P)}-2h(T_n'|T_n,\Psi)\label{ps1}\\
    &\leq 6n\log P \underbrace{-2h(T_n'|T_n,\Psi,W_4)}_{o(\log P)}.
\end{align}
where in (\ref{ps}), $\Psi = \{\Omega^n,W_1,W_2,W_3\}$, and in (\ref{ps1}), $T_n = \{Y_{3,PNNN}^n,Y_{1,NPNN}^n,Y_{1,NNPN}^n,Y_{1,NNNP}^n\}$, $T_n'=\{Y_1^n,Y_3^n\}-T_n$. Therefore, we have
\begin{align}
    2d_1 + 2d_2 + 2d_3 + d_4 &\leq 3\label{e7}
%    d_1 + 2d_2 + 2d_3 + 2d_4 &\leq 3\nonumber\\
%    2d_1 + d_2 + 2d_3 + 2d_4 &\leq 3\nonumber\\
%    2d_1 + 2d_2 + d_3 + 2d_4 &\leq 3\label{e7}.
\end{align}
In the left hand side of (\ref{e7}), user 4 has the coefficient of 1 and the remaining 3 users have the coefficient of 2. Also, instead of changing the roles of user 1 and 3, roles of user 2 and 3 or roles of user 1 and 2 could have been changed. Although this $\binom{3}{2}$ changes would not result in a new inequality due to the structure of the pattern shown in figure \ref{fig101}, these changes of the roles of the remaining 3 users (with coefficient 2) are necessary in general. Therefore, motivated by this simple example, we can have a set of inequalities for the 4-user MISO BC with P and N.
\begin{align}
    2d_1 + 2d_2 + 2d_3 + d_4 &\leq 2 + 4\lambda_P \nonumber\\&\ \ \ + \min\{\lambda_{PP--},\lambda_{P-P-},\lambda_{-PP-}\}\nonumber\\
    d_1 + 2d_2 + 2d_3 + 2d_4 &\leq 2 + 4\lambda_P \nonumber\\&\ \ \ + \min\{\lambda_{-PP-},\lambda_{-P-P},\lambda_{--PP}\}\nonumber\\
    2d_1 + d_2 + 2d_3 + 2d_4 &\leq 2 + 4\lambda_P \nonumber\\&\ \ \ + \min\{\lambda_{P-P-},\lambda_{P--P},\lambda_{--PP}\}\nonumber\\
    2d_1 + 2d_2 + d_3 + 2d_4 &\leq 2 + 4\lambda_P \nonumber\\&\ \ \ + \min\{\lambda_{P--P},\lambda_{PP--},\lambda_{-P-P}\}\label{e77}
\end{align}
where each inequality in (\ref{e77}) is obtained from $\binom{3}{2}$ inequalities each of which with the same left hand side. The general K-user MISO BC can be addressed by using the following definition
\begin{align}\label{deff}
    \lambda(a,b)&=\mbox{ The probability that the CSIT of users }a\mbox{ and }b\mbox{ is P.}\nonumber\\&\ \ \ \ \  a,b\in[1:K]\ ,\ a\neq b
\end{align}
\textbf{Theorem 2}. Let $\pi^j(.)$ be an arbitrary permutation of size $j$ over $[1:K]$. For the K-user symmetric MISO BC with no delayed CSIT\footnote{The assumptions of symmetric scenario and no delayed CSIT are only used for the readability of formulations. It is important to note that the approach in this section can be applied to the general asymmetric scenario including the delayed CSIT (in this case, the delay is enhanced to perfect instantaneous as in subsection \ref{enh}).}, we have
\begin{align}\label{new11}
    2\sum_{i=1}^{j-1}d_{\pi^j(i)}+d_{\pi^j(j)}&\leq 2 + 2(j-2)\lambda_P\nonumber\\&\ \ \ +\min_{a,b\in[1:j-1]:a<b}\{\lambda(\pi^j(a),\pi^j(b))\}\nonumber\\&\ \ \ \forall \pi^j,j\in[3:K].
\end{align}
\textit{Proof}. The proof is a straightforward generalization of the previous examples.
%The same approach could be easily extended to the $K$-user MISO BC which is omitted for brevity. It is obvious that none of the above inequalities can have its right-hand side written in terms of only marginal probabilities. Therefore, in contrast to the two user scenario, marginal probabilities of CSIT are not sufficient for defining the DoF region of the general $K$-user MISO BC, and having the same marginal probabilities does not guarantee the same DoF region.
\section{On the achievability}\label{s55}
In this section, we consider the bounds in (\ref{sym}) for the symmetric scenario.\footnote{The main goal of this section is to show that these bounds can become tight and are not always loose.}
%The outer bound in theorem consists of $2^K-1+\sum_{j=2}^Kj!\left(\begin{array}{c}K\\j\end{array}\right)$ inequalities. For $K=2$, it is observed that the outer bound (even with $M>2$) will be the same as \cite{Tandon} where it was shown to be achievable, regardless of the pattern of CSIT.
For $K\geq 3$, we show that given the marginal probabilities of CSIT, there exists at least one CSIT pattern that achieves the outer bound in the following two scenarios.
\subsection{$\lambda_D = 0$}
\begin{figure}[t]
  \centering
  \includegraphics[width=9cm]{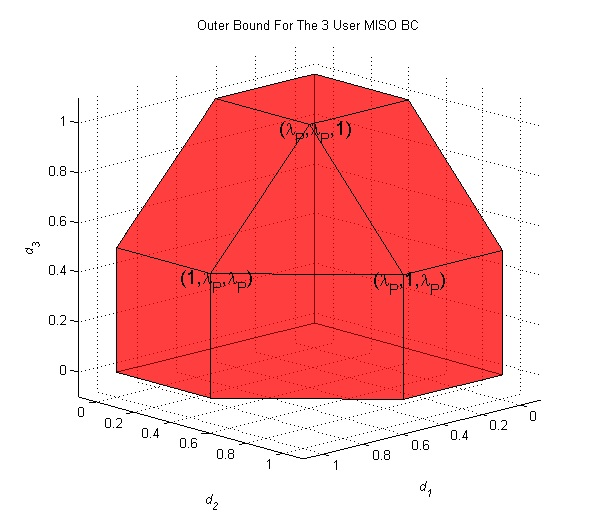}\\
  \caption{Region in case A for 3 user BC}\label{fig1}
\end{figure}

In this case, $2^K-1$ inequalities are active and the remaining inequalities become inactive.
The reason can be easily verified from the inequalities, however, a simpler intuitive way is to consider that when there is no delayed CSIT, those inequalities derived from the degraded broadcast channel are inactive.
In this case, the region is defined by $2^K-1$ hyperplanes in $R_+^K$ and has the following K corner points
\begin{equation}\label{corner}
  (1,\lambda_P,\ldots,\lambda_P),(\lambda_P,1,\lambda_P,\ldots,\lambda_P),\ldots,(\lambda_P,\ldots,\lambda_P,1)
\end{equation}
The corner points have the unique characteristic that the whole region can be constructed by time sharing between them. Therefore, the achievability of these points is equivalent to the achievability of the whole region.
Figure \ref{fig1} shows the region for the 3 user broadcast channel.
The corner points are simply achieved by a scheme that has $N$ time slots and consists of two parts: in the first $\lambda_PN$ time slots, zero forcing beamforming (ZFBF) is carried out where each user receives one interference-free symbol. In the remaining $\lambda_NN$ time slots, only one particular user (depending on the corner point of interest) is scheduled.
\subsection{$\lambda_N\leq \frac{\lambda_D}{\sum_{j=2}^K\frac{1}{j}}$}
Before going further, we need the following simple lemma.

\textbf{Lemma 3}. The minimum probability of delayed CSIT for sending order-$j$ symbols in the $K$-user MAT is
\begin{equation}\label{e9}
  \lambda_D^{min}(K,j)=1-\frac{K-j+1}{K\sum_{i=j}^K\frac{1}{i}}.
\end{equation}
%Substituting $j=1$ in (\ref{e9}), we get the minimum $\lambda_D$ for order-$1$ symbols as
%\begin{equation}
%  \lambda_D^{min}(K)=1-\frac{1}{\sum_{i=j}^K\frac{1}{i}}.
%\end{equation}
%\begin{figure}[t]
%  \centering
%  % Requires \usepackage{graphicx}
%  \includegraphics[width=8cm]{Picture1}\\
%  \caption{Achievable scheme in case A for 3 user BC}\label{fig2}
%\end{figure}
\begin{proof} From \cite{MAT}, the MAT algorithm is based on a concatenation of $K$ phases. Phase $j$ takes $(K-j+1)\binom{K}{j}$ order-$j$ messages as its input, takes $\binom{K}{j}$  time slots and produces $j\binom{K}{j+1}$ order-$j+1$ messages as its output.
%\[(K-j+1)\left(\begin{array}{c}K\\j\end{array}\right) \mbox{order-}j\to \underbrace{{\mbox{Phase }j}}_{\left(\begin{array}{c}K\\j\end{array}\right)\mbox{time slots}}\to j\left(\begin{array}{c}K\\j+1\end{array}\right)\mbox{order-}j+1.\]
In each time slot of phase $j$, the transmitter sends a random linear combination of the $(K-j+1)$ symbols to a subset $S$ of receivers , $|S|=j$. Sending the overheard interferences from the remaining $(K-j)$ receivers to receivers in subset S enables them to successfully decode their $(K-j+1)$ symbols by constructing a set of $(K-j+1)$ linearly independent equations. Therefore, the transmitter needs to know the channel of only $(K-j)$ receivers. In other words, at each time slot of phase $j$, the feedback of $(K-j)$ CSI is enough.
In the MAT algorithm the number of output symbols that phase $j$ produces should match the number of input symbols of phase $j+1$. The ratio between the input of phase $j+1$ and output of phase $j$ is:
\[\frac{(K-j)\binom{K}{j+1}}{j\binom{K}{j+1}}=\frac{(K-j)}{j}.\]
This means that $(K-j)$ repetition of phase $j$ will produce the inputs needed by $j$ repetition of phase $j+1$. In general, in order to have an integer number for repetitions, we multiply phase $1$ by $K!$ (i.e., repeat it $K!$ times), phase $2$ by $\frac{K!}{(K-1)}$, and so on. Therefore, phase $j$ will be repeated $((j-1)!(K-j)!)K$ times which takes $((j-1)!(K-j)!)K\binom{K}{j}$ time slots. Since $(K-j)$ feedbacks from each time slot is sufficient, the number of feedbacks will be $((j-1)!(K-j)!)K\binom{K}{j}(K-j)$. For a successive decoding or order-$j$ symbols, all the higher order symbols must be decoded successfully. Therefore, instead of having delayed CSIT at all time instants from all users, the minimum probability of delayed CSIT is the number of feedbacks from phase $j$ to $K$ divided by the whole number of time slots multiplied by the number of users,
%\[\lambda_D^{min}(K,j)=\frac{\sum_{i=j}^K(i-1)!(K-i)!K\binom{K}{i}(K-i)}
%{\sum_{i=j}^K(i-1)!(K-i)!K\binom{K}{j+1}K}=1-\frac{K-j+1}{K\sum_{i=j}^K\frac{1}{i}}.\] \qedhere
\begin{align*}
    \lambda_D^{min}(K,j)&=\frac{\sum_{i=j}^K(i-1)!(K-i)!K\binom{K}{i}(K-i)}{\sum_{i=j}^K(i-1)!(K-i)!K\binom{K}{j+1}K}\\&=1-\frac{K-j+1}{K\sum_{i=j}^K\frac{1}{i}}.
\end{align*}
\end{proof}
\begin{figure}[t]
  \centering
  \includegraphics[width=9cm]{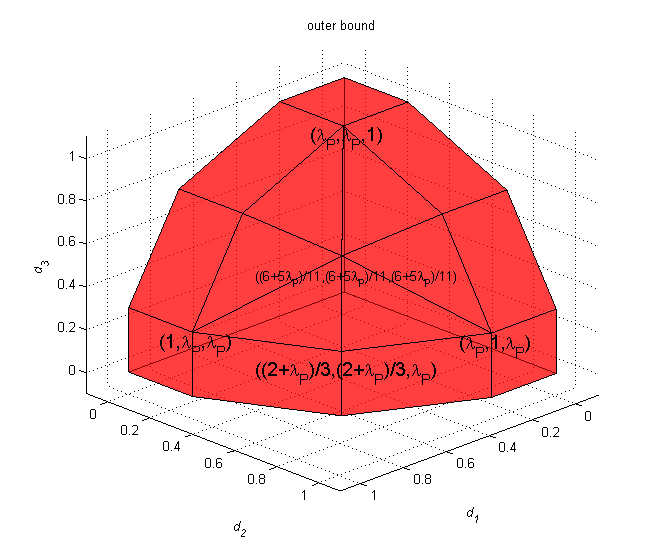}\\
  \caption{Region in case B for 3 user BC}\label{fig3}
\end{figure}
\begin{figure}
  \centering
  \includegraphics[width=8cm]{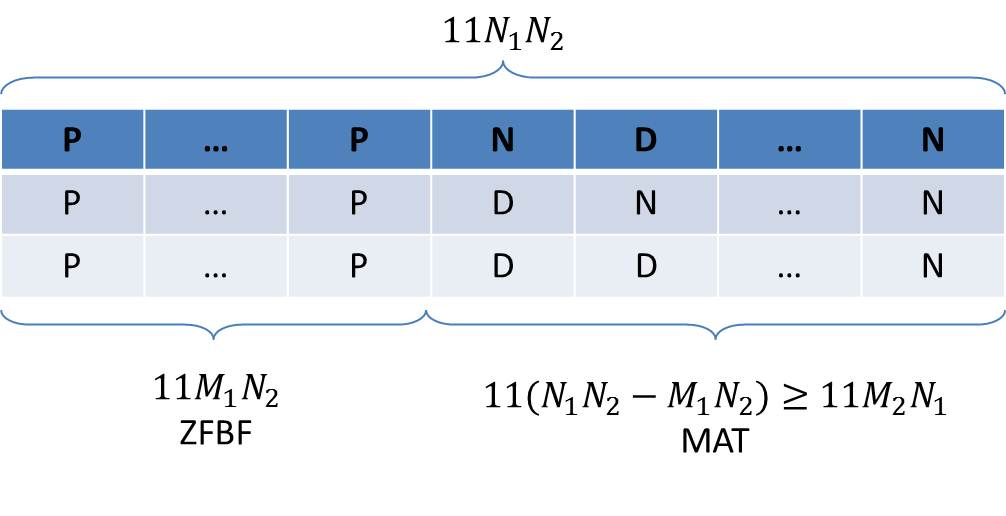}\\
  \caption{Achievable scheme in case B for 3 user BC}\label{fig4}
\end{figure}
In this case (i.e., $\lambda_N\leq \frac{\lambda_D}{\sum_{j=2}^K\frac{1}{j}}$),
the $2^K-K-1$ inequalities having $\sum_i d_i$ (summation with equal weights) in the left-hand side become inactive and the remaining $\sum_{j=1}^Kj!\binom{K}{j}$  inequalities are active which construct $\sum_{j=1}^Kj!\binom{K}{j}$  hyperplanes in $R_+^K$.
The region has $2^K-1$ corner points. In other words, if the coordinates of a point are shown as $(p_1,p_2,\ldots,p_K)$, there are $\binom{K}{j}$ ($j\in[1:K]$) points where $j$ of its $K$ coordinates are $\frac{1+\lambda_P\sum_{i=2}^j\frac{1}{i}}{\sum_{i=1}^j\frac{1}{i}}$ and the remaining $K-j$ coordinates are $\lambda_P$.
%\begin{itemize}
%  \item $\left(\begin{array}{c}K\\1\end{array}\right) \mbox{corner points in the form } (1,\lambda_P,\ldots,\lambda_P),(\lambda_P,1,\lambda_P,\ldots,\lambda_P),\ldots,(\lambda_P,\ldots,\lambda_P,1) $
%  \item $\left(\begin{array}{c}K\\2\end{array}\right) \mbox{corner points in the form } (\frac{2+\lambda_P}{3},\frac{2+\lambda_P}{3},\lambda_P,\ldots,\lambda_P),(\frac{2+\lambda_P}{3},\lambda_P,\frac{2+\lambda_P}{3},\lambda_P,\ldots,\lambda_P),\ldots $
%  \item $\left(\begin{array}{c}K\\3\end{array}\right) \mbox{corner points in the form } (\frac{6+5\lambda_P}{11},\frac{6+5\lambda_P}{11},\frac{6+5\lambda_P}{11},\lambda_P,\ldots,\lambda_P),\ldots $
%  \item \ \ $\ldots, \mbox{and finally,}\left(\begin{array}{c}K\\K\end{array}\right)\mbox{corner points in the form }(\frac{1+\lambda_P\sum_{i=2}^K\frac{1}{i}}{\sum_{i=1}^K\frac{1}{i}},\frac{1+\lambda_P\sum_{i=2}^K\frac{1}{i}}{\sum_{i=1}^K\frac{1}{i}},\ldots,\frac{1+\lambda_P\sum_{i=2}^K\frac{1}{i}}{\sum_{i=1}^K\frac{1}{i}})$
%\end{itemize}
The region for the 3 user broadcast channel and the achievable scheme are shown in figure \ref{fig3} and figure \ref{fig4}, respectively.
The achievable scheme is based on a concatenation of ZFBF and MAT as follows.
%For the first $K$ corner points listed above, the achievability scheme is the same as that in the previous section (i.e., ZFBF + fixed user scheduling).
For the $\binom{K}{j}$ corner points, we write %in the form $(\frac{1+\lambda_P\sum_{i=2}^j\frac{1}{i}}{\sum_{i=1}^j\frac{1}{i}},\frac{1+\lambda_P\sum_{i=2}^j\frac{1}{i}}{\sum_{i=1}^j\frac{1}{i}},\ldots,\lambda_P,\ldots,\lambda_P)$,
\begin{equation}\
  \lambda_P=\frac{M_1}{N_1}, \lambda_D=\frac{M_2}{N_2}, \lambda_D^{min}(j,1)=\frac{m}{n}
\end{equation}
where $m,n,M_i$  and $N_i$ ($i=1,2$) are integers. Making a common denominator between $\lambda_P$ and $\lambda_D$ we have
\begin{equation}
 \lambda_P=\frac{nM_1N_2}{nN_1N_2}, \lambda_D=\frac{nN_1M_2}{nN_1N_2}.
\end{equation}
%\begin{figure}[t]
%  \centering
%  % Requires \usepackage{graphicx}
%  \includegraphics[width=10 cm]{Picture4}\\
%  \caption{Two CSIT patterns. $(a)\  \lambda_P^1=\frac{1}{4},\lambda_P^2=\frac{1}{2} \mbox{ and } \lambda_P^3=1\ (b)\  \lambda_P^1=\frac{1}{4} \mbox{ and } \lambda_P^2=\lambda_P^3=\frac{1}{2}$}\label{fig6}
%\end{figure}
We construct $nN_1N_2$ time slots where the CSIT of each user can be Perfect (P) or Delayed (D) in $nM_1N_2$ or $nN_1M_2$ time slots, respectively. In the first $nM_1N_2$ time slots, ZFBF is carried out. In the remaining $n(N_1N_2-M_1N_2)$ time slots, $j$-user MAT algorithm is done.
At each time slot of the ZFBF part, 1 interference-free symbol is received by each user and in the MAT part, $\frac{n(N_1N_2-M_1N_2)}{1+\frac{1}{2}+\cdots+\frac{1}{j}}$ symbols are sent to each of the users in subset $S$ (with $|S|=j$) where $S$ depends on the corner point of interest. In order to do the MAT algorithm in the second part, the minimum probability of delayed CSIT should be met
\begin{equation}
  nN_1M_2 \geq \lambda_D^{min}(j)n(N_1N_2-M_1N_2)
\end{equation}
Dividing both sides by $nN_1N_2$,
\begin{equation}
  \lambda_D \geq \lambda_D^{min}(j,1)(1-\lambda_P)=\lambda_D^{min}(j,1)(\lambda_D+\lambda_N)
\end{equation}
which results in
\begin{equation}
 \lambda_N \leq \frac{\lambda_D}{\sum_{i=2}^j\frac{1}{i}}.
\end{equation}
Since it should be valid for all $j$, we have
\begin{equation}
 \lambda_N \leq \frac{\lambda_D}{\sum_{i=2}^K\frac{1}{i}}.
\end{equation}
which is the condition assumed in this case.

\begin{figure}[t]
  \centering
  \includegraphics[width=8cm]{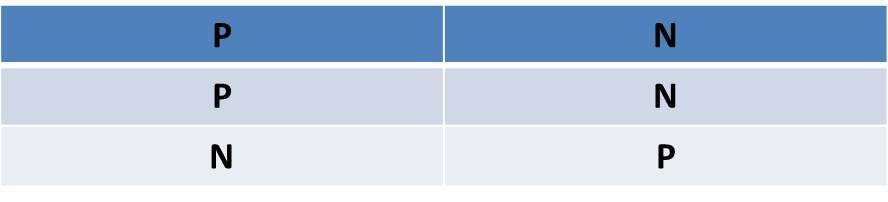}\\
  \caption{An example.}\label{figfig0}
\end{figure}

\begin{figure}[t]
  \centering
  \includegraphics[width=9cm]{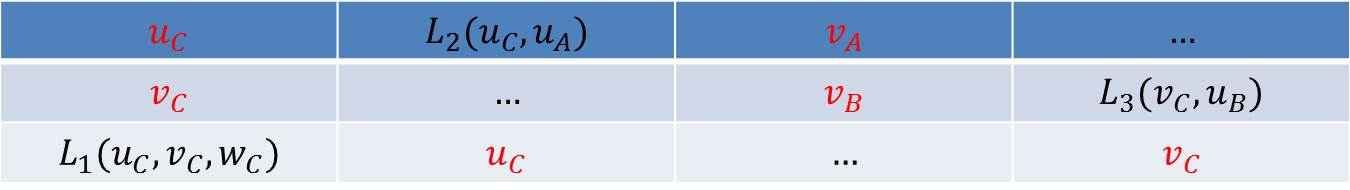}\\
  \caption{The achievable scheme for the boundary point $(\frac{1}{2},\frac{1}{2},\frac{3}{4})$.}\label{figfig}
\end{figure}

Finally, through an example, we show that the bounds in Theorem 2 can be tight. Consider the pattern shown in figure \ref{figfig0}. According to sections \ref{s3} and \ref{sh}, the DoF region has the following outer bound
\begin{align}
    0\leq d_1,d_2,&d_3\leq 1\ \ ,\ \ d_1+d_2\leq\frac{3}{2}\\
    &2d_1+d_2+2d_3\leq 3\label{re1}\\
    &d_1+2d_2+2d_3\leq 3\label{re2}.
\end{align}
The achievable point $(d_1,d_2,d_3)=(\frac{1}{2},\frac{1}{2},\frac{3}{4})$ makes the inequalities in (\ref{re1}) and (\ref{re2}) tight therefore, it is on the boundary of DoF region. This point is achievable as shown in figure \ref{figfig} where the receivers are called A,B and C. Symbols are shown in red where the transmitter has perfect CSIT and those received signals that are not important in the achievability scheme are shown as "...".

\section{Two user MIMO}\label{MIMO}
In previous sections, the K-user MISO BC was considered. The general MIMO BC is more challenging due to the mismatch between the number of receive antennas\footnote{It is important to note that with different number of antennas, as stated in \cite{Xi}, the dimensions of useful signals and interference signals are not the same in contrast to the symmetric case. Furthermore, the users have different capabilities of decoding which must be taken into account in the achievability schemes.}. In this section, we consider a two user MIMO BC where each user is equipped with $N_k$ ($k\in[1:2]$) antennas and a base station with $M(\geq N_1+N_2)$ antennas wishes to send two independent messages $W_1$ and $W_2$ to their corresponding receivers. The received signal of user $k$ is given by
\begin{equation}\label{1e}
  \mathbf{Y}_k(t)=\mathbf{H}_{k}^{H}(t)\mathbf{X}(t) + \mathbf{W}_k(t) \ ,\ k\in[1:2]\ ,\ t\in[1:n]
\end{equation}
where the channel matrices are assumed to be full rank almost surely. We assume that the CSI of a particular user is either instantaneously Perfect (P) or Not known (N) resulting in the four possible states $PP,PN,NP$ and $NN$ with corresponding probabilities $\lambda_{PP},\lambda_{PN},\lambda_{NP}$ and $\lambda_{NN}$. Let $Y_{i,j}$ denote the received signal at the $j^{th}$ antenna of user $i$ ($i\in[1:2],j\in[1:N_i]$).
Without loss of generality, we assume $N_1\geq N_2$. An outer bound on the DoF region is provided in Theorem 3 and its achievability is discussed afterwards.

\textbf{Theorem 3}. An outer bound for the DoF region of the channel in (\ref{1e}) is given by
\begin{align}
    \frac{d_1}{N_1}+\frac{d_2}{N_2}&\leq 1+\lambda_{PP}+\lambda_{NP}=1+\lambda_P^2\label{2e}\\
    d_1+d_2&\leq N_1+N_2(\lambda_{PP}+\lambda_{PN})=N_1+N_2\lambda_P^1\label{3e}
\end{align}
\begin{proof}
By enhancing user 1 with the message of user 2, Fano's inequality (ignoring $n\epsilon_n$) results in
\begin{align}
    nR_1&\leq I(W_1;\mathbf{Y}_1^n|\Omega^n,W_2)\nonumber\\&=h(\mathbf{Y}_1^n|\Omega^n,W_2)-\underbrace{h(\mathbf{Y}_1^n|\Omega^n,W_1,W_2)}_{no(\log P)}\label{4e}\\
    nR_2&\leq I(W_2;\mathbf{Y}_2^n|\Omega^n)=\underbrace{h(\mathbf{Y}_2^n|\Omega^n)}_{\leq nN_2\log P}-h(\mathbf{Y}_2^n|\Omega^n,W_2)\label{5e}.
\end{align}
Ignoring $o(\log P)$, we have
\begin{align}
    &n(N_2R_1+N_1R_2)\nonumber\\&\leq nN_1N_2\log P + N_2h(\mathbf{Y}_1^n|\Omega^n,W_2)-N_1h(\mathbf{Y}_2^n|\Omega^n,W_2)\label{6e}\\
    &\leq nN_1N_2\log P + \sum_{i=1}^{N_1}h(\Psi_{i}^{N_2}(\Gamma_{N_1})|F)-N_1h(\mathbf{Y}_2^n|F)\label{7e}\\
    &=nN_1N_2\log P + \sum_{i=1}^{N_1}\left[h(\Psi_{i}^{N_2}(\Gamma_{N_1})|F)-h(\mathbf{Y}_2^n|F)\right]\label{7.1e}\\
    &\leq nN_1N_2\log P + nN_1N_2(\lambda_{PP}+\lambda_{NP})\log P\label{8e}
\end{align}
where in (\ref{7e}), $F=\{\Omega^n,W_2\}$ and lemma 1 has been applied with $\Gamma_{N_1}$ denoting the $N_1$ elements of $\mathbf{Y}_1^n$ (i.e., $Y^n_{1,[1:N_1]}$) and $m=N_1-N_2$. Applying the same procedure of section \ref{s3} and lemma 2 to each term of the summation in (\ref{7.1e}) results in (\ref{8e}). By dividing both sides of (\ref{8e}) by $n\log P$ and taking the limit $n,P\to\infty$, (\ref{2e}) is obtained.

For the inequality in (\ref{3e}), we have
\begin{align}
nR_1&\leq I(W_1;\mathbf{Y}_1^n|\Omega^n)\nonumber\\&=I(W_1;Y_{1,[1:N_2]}^n|\Omega^n)+I(W_1;Y_{1,[N_2+1:N_1]}^n|\Omega^n,Y_{1,[1:N_2]}^n)\nonumber\\
&=h(Y_{1,[1:N_2]}^n|\Omega^n)-h(Y_{1,[1:N_2]}^n|\Omega^n,W_1)\nonumber\\&\ \ \ +h(Y_{1,[N_2+1:N_1]}^n|\Omega^n,Y_{1,[1:N_2]}^n)\nonumber\\&\ \ \ -h(Y_{1,[N_2+1:N_1]}^n|\Omega^n,Y_{1,[1:N_2]}^n,W_1)\nonumber\\
&\leq \underbrace{h(Y_{1,[1:N_2]}^n)}_{\leq nN_2\log P}-h(Y_{1,[1:N_2]}^n|\Omega^n,W_1)+\underbrace{h(Y_{1,[N_2+1:N_1]}^n)}_{\leq n(N_1-N_2)\log P}\nonumber\\&\ \ \ -\underbrace{h(Y_{1,[N_2+1:N_1]}^n|\Omega^n,Y_{1,[1:N_2]}^n,W_1,W_2)}_{no(\log P)}\label{8.5e}\\
&\leq nN_1\log P -h(Y_{1,[1:N_2]}^n|\Omega^n,W_1)-no(\log P)\label{9e}
\end{align}
where in (\ref{8.5e}), we used the fact that conditioning reduces the entropy. We enhance user 2 with the message of user 1. Therefore,
\begin{align}\label{10e}
   nR_2&\leq I(W_2;\mathbf{Y}_2^n|\Omega^n,W_1)\nonumber\\&=h(\mathbf{Y}_2^n|\Omega^n,W_1)-\underbrace{h(\mathbf{Y}_2^n|\Omega^n,W_1,W_2)}_{no(\log P)}.
\end{align}
By adding (\ref{9e}) and (\ref{10e}), we get
\begin{align}\label{11e}
    nR_1 + nR_2 &\leq \underbrace{h(\mathbf{Y}_2^n|\Omega^n,W_1)-h(Y_{1,[1:N_2]}^n|\Omega^n,W_1)}_{\leq nN_2(\lambda_{PP}+\lambda_{PN})\log P}\nonumber\\&\ \ \ +nN_1\log P - 2no(\log P)
\end{align}
where the same procedure of section \ref{s3} has been applied to the difference in (\ref{11e}). Therefore,
\begin{equation}\label{12e}
    d_1+d_2\leq N_1+N_2(\lambda_{PP}+\lambda_{PN})=N_1+N_2\lambda_P^1.
\end{equation}
\end{proof}
In the sequel, we show that when $\lambda_{PN}\leq \frac{N_2}{N_1}\lambda_{NP}$, the outer bound, which is defined by (\ref{2e}) and (\ref{3e}), is tight. Specifically, we show the achievability of the inner bound defined by the following inequalities
\begin{align}
    \frac{d_1}{N_1}+\frac{d_2}{N_2}&\leq 1+\lambda_P^2\label{13e}\\
    d_1+d_2&\leq N_1+N_2(\lambda_{PP}+\min(\lambda_{PN},\frac{N_2}{N_1}\lambda_{NP}))\label{14e}.
\end{align}
It is obvious that when $\lambda_{PN}\leq \frac{N_2}{N_1}\lambda_{NP}$, the inner bound coincides with the outer bound. We consider a block of $n$ (sufficiently large) time instants. In this block, there are $n\lambda_{PN}$ time instants in the $PN$ state (i.e., where the CSI of user 1 is perfectly known and CSI of user 2 is unknown), $n\lambda_{NP}$ time instants in the $NP$ state, $n\lambda_{PP}$ time instants in the $PP$ state and $n\lambda_{NN}$ time instants in the $NN$ state. Without loss of generality, we assume $n$ is chosen in such a way that all these numbers are integers. From now on, whenever it is said that $N$ symbols are sent orthogonal to the matrix $\mathbf H$, it is meant that these $N$ symbols are precoded by a matrix whose columns are chosen from the null space of $\mathbf{H}^H$.

 The following achievable schemes are based on a simple interference cancellation scheme. In other words, if at each of the $m$ time instants in the $PN$ state, $N_1$ private symbols are sent to user 1 and $N_2$ private symbols are sent (orthogonal to the channel of user 1) to user 2, user 2 needs to get rid of $nN_1$ interfering symbols from user 1 to decode its own symbols. If we pick $m\frac{N_1}{N_2}$ time instants in the $NP$ state, at each of these time instants, $N_2$ interfering symbols can be sent to user 2 and since these interfering symbols are already known at user 1, $N_1$ new private symbols can be sent (orthogonal to the channel of user 2) to user 1. This could be viewed as a generalization of the $S_{3}^{\frac{3}{2}}$ \cite{Tandon} to the MIMO case where the mismatch between the number of receiving antennas across the users is taken into account. The achievability is divided into two scenarios.

\subsection{$N_1\lambda_{PN}\leq N_2\lambda_{NP}$}
In this case, the region is shown in figure \ref{DoF2}.
\subsubsection*{\textbf{A.1}} 
\begin{figure*}[!t]
\normalsize
\centering
  \includegraphics[width=12cm]{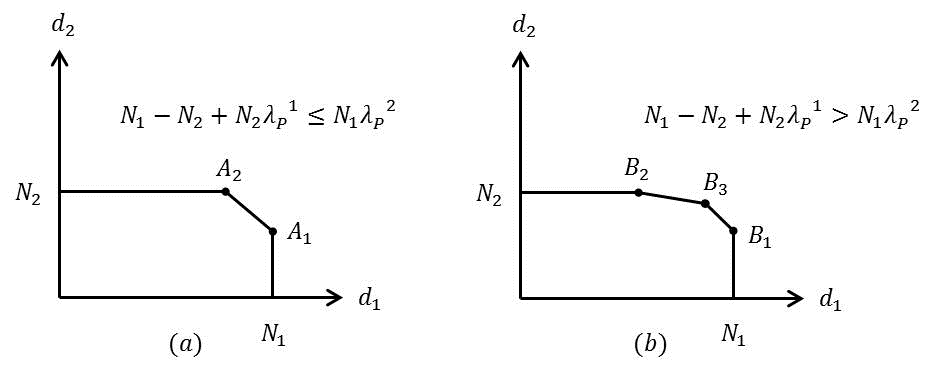}\\
  \caption{The DoF region when $N_1\lambda_{PN}\leq N_2\lambda_{NP}$.}\label{DoF2}
\hrulefill
\vspace*{4pt}
\end{figure*}

When $N_1-N_2+N_2\lambda_P^1\leq N_1\lambda_P^2$, the region (figure \ref{DoF2} (a)) has the corner points $A_1(N_1,N_2\lambda_P^1)$ and $A_2(N_1-N_2+N_2\lambda_P^1,N_2)$.

The achievability of $A_1$ is as follows.

\textbf{Phase 1}: At each of the $n\lambda_{PN}$ time instants, $N_1$ and $N_2$ private symbols are sent to user 1 and user 2, respectively. These $N_2$ private symbols are sent orthogonal to $\mathbf{H}_1(t)$. Therefore, user 1 receives its intended $nN_1\lambda_{PN}$ symbols and user 2 receives $n(N_1+N_2)\lambda_{PN}$ symbols. User 1 can decode its symbols immediately, while user 2 has to get rid of $nN_1\lambda_{PN}$ interfering symbols.
%\begin{figure}[t]
%  \centering
%  % Requires \usepackage{graphicx}
%  \includegraphics[width=10cm]{Fig10}\\
%  \caption{The DoF region when $N_1\lambda_{PN}\leq N_2\lambda_{NP}$.}\label{DoF2}
%\end{figure}
\textbf{Phase 2}: Among the $n\lambda_{NP}$ time instants in the $NP$ state,  $\frac{N_1}{N_2}n\lambda_{PN}(\leq n\lambda_{NP})$ time instants are selected. At each of these selected time instants, $N_2$ interfering symbols of phase 1 are sent to user 2 and $N_1$ new private symbols are sent to user 1. These $N_1$ private symbols are sent orthogonal to $\mathbf{H}_2(t)$. User 2 receives the $nN_1\lambda_{PN}$ interfering symbols which enables it to decode its private symbols in phase 1. The interfering symbols of user 2 are already known at user 1, therefore, user 1 can successfully decode its private symbols in this phase.

\textbf{Phase 3}: In the remaining time instants in the $NP$ state (i.e., $n\lambda_{NP}-\frac{N_1}{N_2}n\lambda_{PN}$ ) and all the $n\lambda_{NN}$ time instants, $N_1$ private symbols are sent to user 1.

\textbf{Phase 4}: In all the $n\lambda_{PP}$ time instants, $N_1$ and $N_2$ private messages orthogonal to $\mathbf H_2(t)$ and $\mathbf H_1(t)$, respectively are sent to user 1 and user 2.

Therefore, user 1 and user 2 can, respectively, decode $nN_1$ and $nN_2(\lambda_{PP}+\lambda_{PN})$ private symbols in the block of $n$ time instants which achieves the first corner point ($N_1,N_2\lambda_P^1$).

The achievability of $A_2$ is as follows.

\textbf{Phase 1}: Among the $n\lambda_{NP}$ time instants in the $NP$ state, $n\frac{N_1}{N_2}\lambda_{PN}$ time instants are selected. At each of these selected time instants, $N_2$ and $N_1$ private symbols are sent to user 2 and user 1, respectively. These $N_1$ private symbols are sent orthogonal to $\mathbf{H}_2(t)$.  Therefore, user 2 can decode $nN_1\lambda_{PN}$ private symbols and user 1 receives $n(N_1+N_2)\frac{N_1}{N_2}\lambda_{PN}$ symbols of which $nN_1\lambda_{PN}$ symbols are interferers.

\textbf{Phase 2}: At each of the $n\lambda_{PN}$ time instants, $N_1$ interfering symbols in phase 1 are sent to user 1 and $N_2$ private symbols to user 2. These $N_2$ private symbols are sent orthogonal to $\mathbf{H}_1(t)$. Therefore, user 1 is able to decode its private symbols in phase 1.

\textbf{Phase 3}: There are $n\lambda_{NP}-\frac{N_1}{N_2}n\lambda_{PN}$ remaining time instants in the $NP$ state. $n\lambda_{NN}\frac{(N_1-N_2)}{N_2}$ of them are selected (note that $n\lambda_{NN}\frac{(N_1-N_2)}{N_2}\leq n\lambda_{NP}-\frac{N_1}{N_2}n\lambda_{PN}$ due to the condition in the figure \ref{DoF2}(a)). At each of these selected time instants, $N_2$ and $N_1$ private symbols are sent to user 2 and user 1, respectively. These $N_1$ private symbols are sent orthogonal to $\mathbf{H}_2(t)$. Therefore, user 1 has to get rid of $n\lambda_{NN}(N_1-N_2)$ interfering symbols.

\textbf{Phase 4}: At each of the $n\lambda_{NN}$ time instants, $N_2$ private symbols are sent to user 2 and $N_1-N_2$ interfering symbol from phase 3 are sent to user 1. The interfering symbols are already known at user 2, therefore user 2 successfully decodes its symbols. User 1, having $N_1$ antennas, is capable of decoding all the sent symbols in this phase.

\textbf{Phase 5}: In the remaining time instants in the $NP$ states, $N_2$ and $N_1-N_2$ private symbols are sent to user 2 and user 1, respectively. These $N_1-N_2$ private symbols are sent orthogonal to $\mathbf{H}_2(t)$.

\textbf{Phase 6}: The same as phase 4 for the achievability of $A_1$

Therefore, user 1 and user 2 can, respectively, decode $n(N_1-N_2+N_2\lambda_P^1)$ and $nN_2$ private symbols in the block of $n$ time instants which achieves the second corner point.

\subsubsection*{\textbf{A.2}} When $N_1-N_2+N_2\lambda_P^1> N_1\lambda_P^2$, the region (figure \ref{DoF2} (b)) has the corner points $B_1(N_1,N_2\lambda_P^1)$, $B_2(N_1\lambda_P^2,N_2)$ and $B_3(N_1-\frac{N_1N_2(\lambda_P^2-\lambda_P^1)}{N_1-N_2},\frac{N_1N_2\lambda_P^2-N_2^2\lambda_P^1}{N_1-N_2})$.

The achievability of $B_1$ is the same as that of $A_1$ and the achievability of $B_2$ is as follows.

\textbf{Phase 1 and 2}: Similar to the phase 1 and phase 2 in the achievability of $A_2$.
%Among the $n\lambda_{NP}$ time instants in the $NP$ state, $n\frac{N_1}{N_2}\lambda_{PN}(\leq n\lambda_{NP})$ time instants are selected. At each of these selected time instants, $N_2$ and $N_1$ private symbols are sent to user 2 and user 1, respectively. The $N_1$ private symbols are precoded by a matrix where its columns are chosen from the null space of $\mathbf{H}_2^H$. Therefore, user 2 decodes $nN_1\lambda_{PN}$ private symbols and user 1 receives $n(N_1+N_2)\frac{N_1}{N_2}\lambda_{PN}$ symbols of which $nN_1\lambda_{PN}$ symbols are interferers.
%
%\textbf{Phase 2}: At each of the $n\lambda_{PN}$ time instants, $N_1$ interfering symbols in phase 1 are sent to user 1 and $N_2$ private symbols to user 2. The $N_2$ private symbols are precoded by a matrix where its columns are chosen from the null space of $\mathbf{H}_1^H$. Therefore, user 1 can decode its private symbols in phase 1.

\textbf{Phase 3}: There are $n\lambda_{NP}-\frac{N_1}{N_2}n\lambda_{PN}$ remaining time instants in the $NP$ state. At each of these remaining time instants, $N_2$ and $N_1$ private symbols are sent to user 2 and user 1, respectively. These $N_1$ private symbols are sent orthogonal to $\mathbf{H}_2(t)$. Therefore, user 1 has to get rid of $nN_2\lambda_{NP}-nN_1\lambda_{PN}$ interfering symbols.

\textbf{Phase 4}: There are $n\lambda_{NN}$ time instant in the $NN$ state. $\frac{nN_2\lambda_{NP}-nN_1\lambda_{PN}}{N_1-N_2}$ of them are selected (note that $n\lambda_{NN}> \frac{nN_2\lambda_{NP}-nN_1\lambda_{PN}}{N_1-N_2}$ due to the condition in the figure \ref{DoF2}(b)) At each of these selected time instants, $N_2$ private symbols are sent to user 2 and $N_1-N_2$ interfering symbols from phase 3 are sent to user 1. Therefore, with $N_1$ antennas, user 1 can decode its private symbols in phase 3. Since these interfering symbols are already known at user 2, it can successfully decode its $N_2$ private symbols in this phase.

\textbf{Phase 5}: In the remaining time instants in the $NN$ states, $N_2$ private symbols are sent to user 2.

\textbf{Phase 6}: The same as phase 4 in the achievability $A_1$.

Therefore, user 1 and user 2 can, respectively, decode $nN_1\lambda_P^2$ and $nN_2$ private symbols in the block of $n$ time instants which achieves the second corner point.

The achievability of $B_3$ follows the same lines as the achievability of $B_2$ except that in phase 5, in the remaining $NN$ time instants, instead of sending $N_2$ private symbols to user 2, $N_1$ private symbols are sent to user 1.

In conclusion, the outer bound in Theorem 3 is the optimal DoF region in this case (i.e., $N_1\lambda_{PN}\leq N_2\lambda_{NP}$).
\subsection{$N_1\lambda_{PN}> N_2\lambda_{NP}$}
\begin{figure*}[!t]
\normalsize
\centering
  \includegraphics[width=12cm]{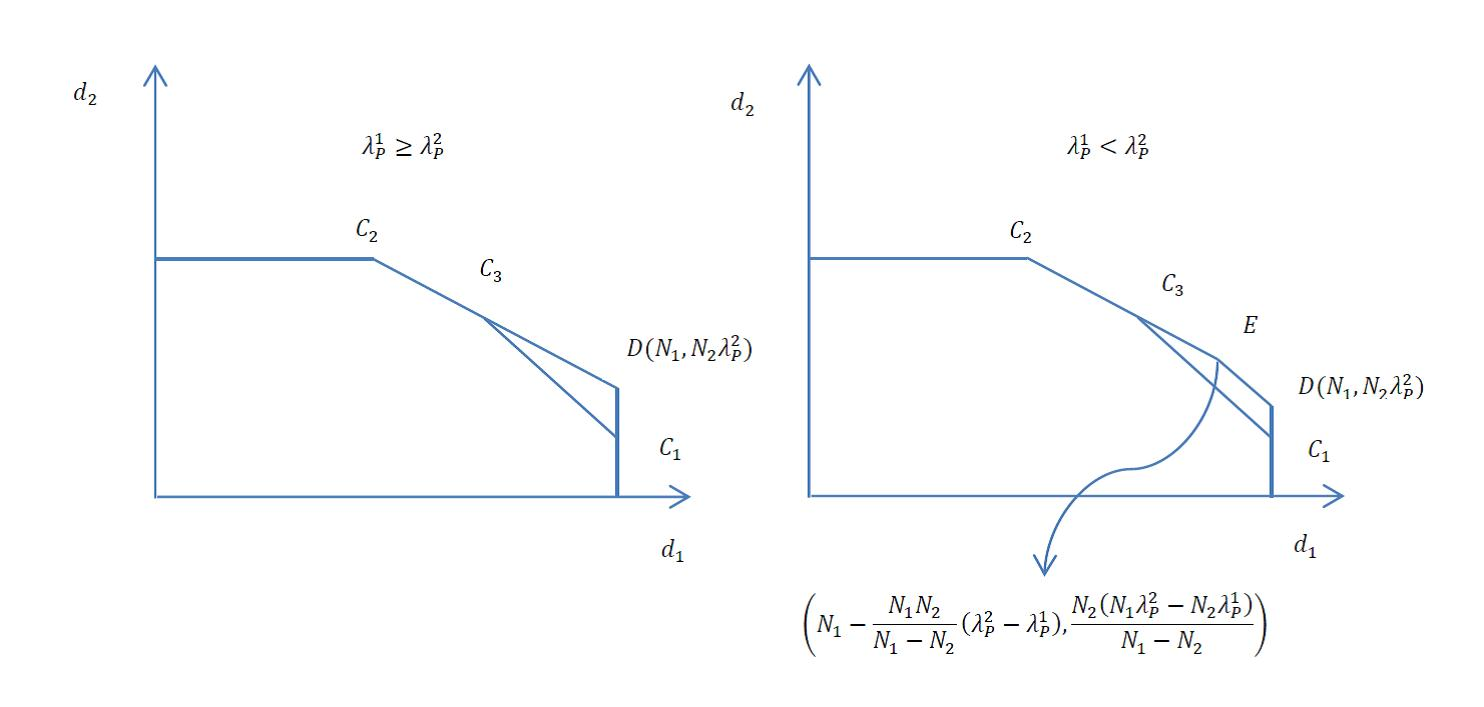}\\
  \caption{The achievable DoF region (i.e., the inner bound) and the outer bound when $N_1\lambda_{PN}> N_2\lambda_{NP}$.}\label{Compar}
\hrulefill
\vspace*{4pt}
\end{figure*}

In this case, the achievable region has three corner points $C_1(N_1,N_2\lambda_{PP}+\frac{N_2^2}{N_1}\lambda_{NP})$, $C_2(N_1\lambda_P^2,N_2)$ and $C_3(N_1-N_2\lambda_{NP},N_2\lambda_P^2+\frac{N_2^2}{N_1}\lambda_{NP})$. This is shown in figure \ref{Compar} along with the outer bound where the outer bound has two corner points ($C_2$ and $D$) when $\lambda_P^1\geq\lambda_P^2$ and three corner points otherwise (i.e., $C_2$, $E$ and $D$).
%\begin{figure}[t]
%  \centering
%  % Requires \usepackage{graphicx}
%  \includegraphics[width=12cm]{Fig11}\\
%  \caption{The achievable DoF region (i.e., the inner bound) and the outer bound when $N_1\lambda_{PN}> N_2\lambda_{NP}$.}\label{Compar}
%\end{figure}

The achievability of $C_1$ is as follows.

\textbf{Phase 1}: There are $n\lambda_{PN}$ time instants in the $PN$ state and $n\frac{N_2}{N_1}\lambda_{NP}(\leq n\lambda_{PN})$ of them are selected. At each of these selected time instants, $N_1$ and $N_2$ private symbols are sent to user 1 and user 2, respectively. These $N_2$ symbols are sent orthogonal to $\mathbf{H}_1(t)$. Therefore, user 1 receives its intended $nN_2\lambda_{NP}$ symbols and user 2 receives $n(N_1+N_2)\frac{N_2}{N_1}\lambda_{NP}$ symbols. User 1 can decode its symbols immediately, while user 2 has to get rid of $nN_2\lambda_{NP}$ interfering symbols.

\textbf{Phase 2}: At each of the $n\lambda_{NP}$ time instants in the $NP$ state, $N_2$ interfering symbols of phase 1 are sent to user 2 and $N_1$ private symbols are sent to user 1. These $N_1$ symbols are sent orthogonal to $\mathbf{H}_2(t)$. User 2 receives the $nN_2\lambda_{NP}$ interfering symbols which enables it to decode its private symbols in phase 1. Since these interfering symbols are already known at user 1, it can successfully decode its $N_1$ private symbols in this phase.

\textbf{Phase 3}: In the remaining time instants in the $PN$ state (i.e., $n\lambda_{PN}-\frac{N_2}{N_1}n\lambda_{NP}$ ) and all the $n\lambda_{NN}$ time instants, $N_1$ private symbols are sent to user 1.

\textbf{Phase 4}: The same as phase 4 in the achievability of $A_1$.

Therefore, user 1 and user 2 can, respectively, decode $nN_1$ and $n(N_2\lambda_{PP}+\frac{N_2^2}{N_1}\lambda_{NP})$ private symbols in the block of $n$ time instants which achieves the first corner point.

The achievability of $C_2$ is as follows.

\textbf{Phase 1}: At each of $n\lambda_{NP}$ time instants, $N_2$ and $N_1$ private symbols are sent to user 2 and user 1, respectively. These $N_1$ symbols are sent orthogonal to $\mathbf{H}_2(t)$. Therefore, user 2 can decode its intended $nN_2\lambda_{NP}$ symbols and user 1 receives $n(N_1+N_2)\lambda_{NP}$ symbols of which $nN_2\lambda_{NP}$ are interferes.

\textbf{Phase 2}: Among the $n\lambda_{PN}$ time instants in the $PN$ state, $n\frac{N_2}{N_1}\lambda_{NP}(\leq n\lambda_{PN})$ time instants are selected. At each of these selected time instants, $N_1$ interfering symbols of phase 1 are sent to user 2 and $N_2$ private symbols are sent to user 2. These $N_2$ symbols are sent orthogonal to $\mathbf{H}_1(t)$. Therefore, user 1 can decode its private symbols in phase 1.

\textbf{Phase 3}: In the remaining time instants in the $PN$ state (i.e., $n\lambda_{PN}-\frac{N_2}{N_1}n\lambda_{NP}$ ) and all the $n\lambda_{NN}$ time instants, $N_2$ private symbols are sent to user 2.

\textbf{Phase 4}: The same as phase 4 in the achievability of $A_1$.

Therefore, user 1 and user 2 can, respectively, decode $nN_1\lambda_P^2$ and $nN_2$ private symbols in the block of $n$ time instants which achieves the second corner point.

The achievability of $C_3$ follows the same lines as the achievability of $C_2$ with the difference that in phase 3, in the remaining time instants in the $PN$ state and all the $n\lambda_{NN}$ time instants, instead of sending $N_2$ private symbols to user 2, $N_1$ private symbols are sent to user 1.

\begin{figure}[H]
  \centering
  \includegraphics[width=8.5cm]{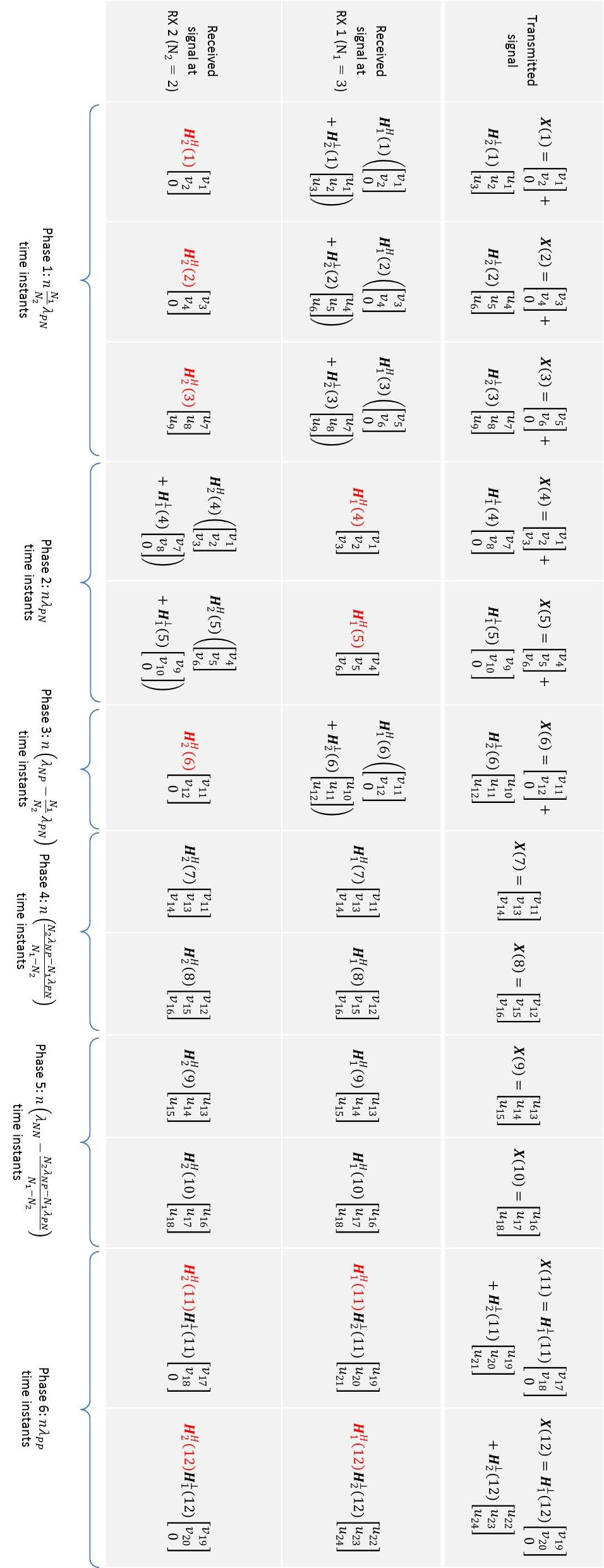}\\
  \caption{An example for achieving the corner point $B_3(N_1-\frac{N_1N_2(\lambda_P^2-\lambda_P^1)}{N_1-N_2},\frac{N_1N_2\lambda_P^2-N_2^2\lambda_P^1}{N_1-N_2})=(2,\frac{5}{3})$. In this example $\lambda_{PN}=\lambda_{PP}=\frac{1}{6}, \lambda_{NP}=\lambda_{NN}=\frac{1}{3}$.}\label{DoF3}
\end{figure}

As an example, figure \ref{DoF3} shows the achievability of the corner point $B_3$ in figure \ref{DoF2}(b). In this example, $\lambda_{PN}=\lambda_{PP}=\frac{1}{6}, \lambda_{NP}=\lambda_{NN}=\frac{1}{3}$. $u$ and $v$ are private symbols from (independently) Gaussian encoded codewords for user 1 and user 2, respectively and $n = 12$. When the CSI of a user is known at the transmitter, it is shown in red.
\section{Conclusion}\label{s7}
Given the marginal probabilities of CSIT, an outer bound was derived for the DoF region of the $K$-user MISO BC with alternating/hybrid CSIT . This outer bound was shown to be achievable by specific CSIT patterns in certain regions. A set of inequalities was provided based on the joint CSIT distribution which shows that in general, the DoF region of the $K$-user MISO BC (when $K\geq 3$) cannot be characterized completely by the marginal probabilities. Finally, an outer bound for the DoF region of a two user MIMO BC in which the CSIT of a user is either perfect or unknown was derived which was shown to be tight in some scenarios.

%------------------------------------------------------------
\appendices
\section{An alternative proof of $\sum_{i=1}^K\frac{d_i}{i}\leq 1 + \sum_{i=2}^K\frac{\sum_{r=1}^{i-1}\lambda_P^r}{i(i-1)}$}\label{s4}
The proof is based on the approach used in \cite{Gesbert}, therefore the following definitions are necessary. The channel vector of user $k$ at time instant $t$ can be written as
\begin{equation}
\mathbf{H}_k(t)=\widehat{\mathbf{H}}_k(t) + \widetilde{\mathbf{H}}_k(t)
\end{equation}
where $\widehat{\mathbf{H}}_k(t)$ and $\widetilde{\mathbf{H}}_k(t)$ are the estimate of the channel and estimation error with distributions $CN({\textbf{0}},(1-\sigma_{k}^{2}(t))\mathbf{I})$ and $CN({\textbf{0}},\sigma_{k}^{2}(t)\mathbf{I})$, respectively.  The variance of error is

\[ \sigma_{k}^{2}(t) = E\left[\|\widetilde{\mathbf{H}}_k(t)\|^{2}\right]. \]
As observed from the above, although the channel is assumed stationary, the estimate is a non-stationary process meaning that the quality of estimation varies over time. The quality of CSIT for user $k$ at time instant $t$ is
\begin{equation}
\alpha_k(t)=-\lim_{P \to \infty}\frac{\log{\left(\sigma_{k}^{2}(t)\right)}}{\log{P}}.
\end{equation}
From the results of \cite{Jindal}, if the rate of feedback scales linearly with $\log P$ (or equivalently, the variance of estimation error decrease as $o(P^{-1})$ or faster), perfect CSIT multiplexing gain can be obtained. Therefore, the effective range of $\alpha_k(t)$ will be $[0,1]$ where in terms of DoF, $\alpha_k(t)=1$ could be interpreted as perfect CSIT of user $k$ at time instant $t$. We also define $\hat\Omega^t$ as the set of all channel estimates up to time instant $t$. Again, for simplicity, we show the inequalities for a fixed permutation of the users while the results could be easily extended to any arbitrary permutations.
As in part \emph{A} of the first proof, the same channel improvement is done here. The only difference is that we assume the users not only have perfect global CSIR, but also they know the channel estimates at the transmitter.
From the chain rule of entropies, each of the terms in the summation in (\ref{e1}) can be written as
\begin{align}\label{em}
\sum_{t=1}^n&\left[\frac{h(Y_{[1:i]}(t)|W_{[1:i-1]},Y_{[1:i]}^{t-1},\Omega^t,\hat{\Omega}^t)}{i}\right.\nonumber\\&\left.-\frac{h(Y_{[1:i-1]}(t)|W_{[1:i-1]},Y_{[1:i-1]}^{t-1},\Omega^t,\hat{\Omega}^t)}{i-1}\right].
\end{align}
%where $Y_i^{t-1}$ is the time extension of $Y$ from time instant $i$ to $t-1$.
By adding $Y_i^{t-1}$ to the conditions of the second entropy, (\ref{em})  will be increased. Therefore,
%\begin{equation}
%\sum_{t=1}^n\left[\frac{h(Y_{[1:i]}(t)|T_{i,t},\Omega(t))}{i}-\frac{h(Y_{[1:i-1]}(t)|T_{i,t},\Omega(t))}{i-1}\right]
%\end{equation}
%\begin{equation}\label{e30}
%    \sum_{i=1}^K\frac{nR_i}{i}\leq \underbrace{h(Y_1^n|\Omega^n)}_{\leq n\log P}+\sum_{i=2}^K\sum_{t=1}^n\left[\frac{h(Y_{[1:i]}(t)|U_{i,t},\Omega(t))}{i}-\frac{h(Y_{[1:i-1]}(t)|U_{i,t},\Omega(t))}{i-1}\right]+no(\log P)
%\end{equation}
%where $U_{i,t}=(W_{[1:i-1]},Y_{[1:i]}^{t-1},\Omega^{t-1})$ and $\Omega(t)$ is the global CSIR at time instant $t$.
\begin{align}\label{ee30}
\sum_{i=1}^K\frac{nR_i}{i}  &\leq\overbrace{h(Y_1^n|\Omega^n,\hat{\Omega}^n)}^{\leq n\log P}\nonumber\\&\ \ \  +\sum_{i=2}^K\sum_{t=1}^n\left[\frac{h(Y_{[1:i]}(t)|U_{i,t},\Omega(t))}{i}\right.\nonumber\\&\ \ \ \left.-
 \frac{h(Y_{[1:i-1]}(t)|U_{i,t},\Omega(t))}{i-1}\right] + no(\log P).
\end{align}
where $U_{i,t}=(W_{[1:i-1]},Y_{[1:i]}^{t-1},\Omega^{t-1},\hat{\Omega}^t)$ and $\Omega(t)$ is the global CSIR at time instant $t$.
In what follows, we find an upper bound for the term in the brackets of (\ref{ee30}). Following the same approach as in \cite{Gesbert}, (\ref{avval}) to (\ref{eee}) are obtained in which we have the Markov chain $\mathbf{X}(t)\leftrightarrow U_{i,t}\leftrightarrow\hat{\Omega}(t)\leftrightarrow \Omega(t)$.
\begin{figure*}[!t]
\normalsize
\begin{align}
   &\max_{P_{U_{i,t}}P_{\mathbf{X}(t)|U_{i,t}}}\left[\frac{h(Y_{[1:i]}(t)|U_{i,t},\Omega(t))}{i}- \frac{h(Y_{[1:i-1]}(t)|U_{i,t},\Omega(t))}{i-1}\right]\label{avval} \\
   &\leq \max_{P_{U_{i,t}}}E_{{U_{i,t}}}\left[\max_{P_{\mathbf{X}(t)|U_{i,t}}}\left(\frac{h(Y_{[1:i]}(t)|U_{i,t}=U,\Omega(t))}{i}-
   \frac{h(Y_{[1:i-1]}(t)|U_{i,t}=U,\Omega(t))}{i-1}\right)\right]\\
   &= \max_{P_{U_{i,t}}}E_{{U_{i,t}}}\left[\max_{P_{\mathbf{X}(t)|U_{i,t}}}E_{\Omega(t)|{U_{i,t}}}\left(\frac{h(Y_{[1:i]}(t)|U_{i,t}=U,\Omega(t)=\mathbf H)}{i}-
   \frac{h(Y_{[1:i-1]}(t)|U_{i,t}=U,\Omega(t)=\mathbf H)}{i-1}\right)\right]\\
      &= \max_{P_{U_{i,t}}}E_{{U_{i,t}}}\left[\max_{P_{\mathbf{X}(t)|U_{i,t}}}E_{\Omega(t)|\hat{\Omega}(t)}\left(\frac{h(\mathbf H_{[1:i]}(t)\mathbf{X}(t)+\mathbf{W}_{[1:i]}(t)|U_{i,t}=U)}{i}-
   \frac{h(\mathbf H_{[1:i-1]}(t)\mathbf{X}(t)+\mathbf{W}_{[1:i-1]}(t)|U_{i,t}=U)}{i-1}\right)\right]\label{equivalent}\\
   &= \max_{P_{U_{i,t}}}E_{{U_{i,t}}}\left[\max_{\textbf{\textit{C}}:\textbf{\textit{C}}\succeq 0,tr(\textbf{\textit{C}})\leq P}\!\!\!\max_{\substack{P_{\mathbf{X}(t)|U_{i,t}}\\Cov(\mathbf{X}(t)|U_{i,t})\preceq \textit{\textbf{C}}}}\!\!\!E_{\Omega(t)|\hat{\Omega}(t)}\left(\frac{h(\mathbf H_{[1:i]}(t)\mathbf{X}(t)+\mathbf{W}_{[1:i]}(t)|U)}{i}-
   \frac{h(\mathbf H_{[1:i-1]}(t)\mathbf{X}(t)+\mathbf{W}_{[1:i-1]}(t)|U)}{i-1}\right)\right]\\
   &= \max_{P_{U_{i,t}}}E_{{U_{i,t}}}\left[\max_{\textbf{\textit{C}}:\textbf{\textit{C}}\succeq0,tr(\textbf{\textit{C}})\leq P}E_{\Omega(t)|\hat{\Omega}(t)}\left(\frac{\log\det{(I_i+\mathbf H_{[1:i]}(t)\textbf{K}_*\mathbf H^H_{[1:i]}(t))}}{i}-
   \frac{\log\det{(I_{i-1}+\mathbf H_{[1:i-1]}(t)\textbf{K}_*\mathbf H^H_{[1:i-1]}(t))}}{i-1}\right)\right] \label{ee1}\\
   &\leq E_{\hat{\Omega}(t)}\left[\max_{\textbf{\textit{K}}:\textbf{\textit{K}}\succeq0,tr(\textbf{\textit{K}})\leq P} E_{\Omega(t)|\hat{\Omega}(t)} \left(\frac{\log\det{(I_i+\mathbf H_{[1:i]}(t)\textbf{K}\mathbf H^H_{[1:i]}(t))}}{i}
  -\frac{\log\det{(I_{i-1}+\mathbf H_{[1:i-1]}(t)\textbf{K}\mathbf H^H_{[1:i-1]}(t))}}{i-1}\right)\right] \\
   &\leq  -\frac{\log\det{(\Sigma^2)}}{i(i-1)}+o(\log P) \label{eee}
\end{align}
\hrulefill
\vspace*{4pt}
\end{figure*}
In (\ref{equivalent}), we have written the signals of the users in terms of their concatenated channels and noise terms. (\ref{ee1}) is the application of extremal inequality \cite{extremal}, \cite{weingarten} where the Gaussian distribution maximizes a specific difference between two differential entropies. The last inequality (\ref{eee}) comes from (101) in \cite{Elia}, in which
\begin{equation*}
  \Sigma^2=\mbox{diag}\left(\sigma_{[1:i-1]}^2(t)\right).
\end{equation*}
Therefore, we can write
\begin{align}
 \sum_{i=1}^K\frac{nR_i}{i}&\leq n\log P +
 \sum_{i=2}^K\sum_{t=1}^n\left[-\frac{\log\det{(\Sigma^2)}}{i(i-1)}+o(\log P)\right ]\nonumber\\&\ \ \ +no(\log P) \nonumber \\
 &= n\log P + \sum_{i=2}^K\frac{\sum_{t=1}^n[\alpha_1(t)+\cdots+\alpha_{i-1}(t)]}{i(i-1)}\log P
 \nonumber\\&\ \ \ + nKo(\log P).
\end{align}
Since the channel is degraded and $D$ is replaced with $N$, the CSIT is either $P$ or $N$. Therefore,  the $\alpha$'s are either $1$ with probability $\lambda_P^i$ or $0$ otherwise. Hence, for $n$ large enough, we have
\begin{equation*}
  \lim_{n\to \infty}\sum_{t=1}^n[\alpha_1(t)+\cdots+\alpha_{i-1}(t)]=n\sum_{r=1}^{i-1}\lambda_P^r
\end{equation*}
which results in
\begin{equation}\label{e2}
  \sum_{i=1}^K\frac{nR_i}{i}\leq n\log P + \sum_{i=2}^K\sum_{r=1}^{i-1}\frac{n\lambda_P^r}{i(i-1)}\log P
  + nKo(\log P)
\end{equation}
at large $n$. Dividing both sides by $n\log P$ and taking the limit of (\ref{e2}) as $n,P\to \infty$, we get
\begin{equation}
 \sum_{i=1}^K\frac{d_i}{i}\leq 1 + \sum_{i=2}^K\frac{\sum_{r=1}^{i-1}\lambda_P^r}{i(i-1)}.
\end{equation}
It is obvious that the same approach can be applied to any other permutations of $(1,2,\ldots,K)$.

\bibliography{REFERENCE}
\bibliographystyle{IEEEtran}
\begin{IEEEbiographynophoto}{Borzoo Rassouli}
received his M.Sc. degree in communication systems engineering from University of Tehran, Iran, in 2012. He is currently pursuing the Ph.D. degree in the Communication and Signal Processing Group, Imperial College, London, UK. His research interests are in the general area of information theory, wireless communications, detection and estimation theory.
\end{IEEEbiographynophoto}

\begin{IEEEbiographynophoto}{Chenxi Hao}
received the B.Sc. and M.Sc. degrees from Beijing University of Posts and Telecommunications in 2010 and University of Southampton in 2011,
 respectively. He is currently pursuing the Ph.D. degree in the Communication and Signal Processing Group, Imperial College London. He is also with Beijing Samsung Telecom R\&D Center. His current research interests lie in communication theory, network information theory and MIMO systems with limited feedback.
\end{IEEEbiographynophoto}

\begin{IEEEbiographynophoto}{Bruno Clerckx}
received the M.S. and Ph.D. degrees
in applied science from Universite catholique de
Louvain, Belgium. He is now a Senior Lecturer (Associate
Professor) at Imperial College London. He held visiting
research positions at Stanford University and
EURECOM and was with Samsung Electronics from
2006 to 2011. He actively contributed to 3GPP LTE/
LTE-A and IEEE802.16m.
He is the author or coauthor of two books on
MIMO wireless communications and networks and
numerous research papers, standard contributions
and patents. He received the Best Student Paper Award at the IEEE Symposium
on Communications and Vehicular Technology in 2002 and several awards from
Samsung in recognition of special achievements. Dr. Clerckx has served as an
Editor for IEEE TRANSACTIONS ON COMMUNICATIONS and is currently an Editor for IEEE TRANSACTIONS ON WIRELESS COMMUNICATIONS.
\end{IEEEbiographynophoto}
\end{document}